\documentclass[sigconf]{acmart}

\usepackage{multirow}
\usepackage{lscape}

\AtBeginDocument{%
  }

\copyrightyear{2025}
\acmYear{2025}
\setcopyright{cc}
\setcctype{by}
\acmConference[CHI '25]{CHI Conference on Human Factors in Computing Systems}{April 26-May 1, 2025}{Yokohama, Japan}
\acmBooktitle{CHI Conference on Human Factors in Computing Systems (CHI '25), April 26-May 1, 2025, Yokohama, Japan}\acmDOI{10.1145/3706598.3714226}
\acmISBN{979-8-4007-1394-1/25/04}
\acmDOI{10.1145/3706598.3714226}

\begin{document}

\title[Understanding Players' In-game Assessment of Communication Processes in League of Legends]{Less Talk, More Trust: Understanding Players' In-game Assessment of Communication Processes in League of Legends}

\author{Juhoon Lee}
\email{juhoonlee@kaist.ac.kr}
\affiliation{%
  \institution{KAIST}
  \city{Daejeon}
  \country{Republic of Korea}
}

\author{Seoyoung Kim}
\email{youthskim@kaist.ac.kr}
\affiliation{%
  \institution{KAIST}
  \city{Daejeon}
  \country{Republic of Korea}
}

\author{Yeon Su Park}
\email{yeonsupark@kaist.ac.kr}
\affiliation{%
  \institution{KAIST}
  \city{Daejeon}
  \country{Republic of Korea}
}

\author{Juho Kim}
\email{juhokim@kaist.ac.kr}
\affiliation{%
  \institution{KAIST}
  \city{Daejeon}
  \country{Republic of Korea}
}

\author{Jeong-woo Jang}
\email{jangjw29@kaist.ac.kr}
\affiliation{%
  \institution{KAIST}
  \city{Daejeon}
  \country{Republic of Korea}
}

\author{Joseph Seering}
\email{seering@kaist.ac.kr}
\affiliation{%
  \institution{KAIST}
  \city{Daejeon}
  \country{Republic of Korea}
}

\renewcommand{\shortauthors}{Lee et al.}

\begin{abstract}
  In-game team communication in online multiplayer games has shown the potential to foster efficient collaboration and positive social interactions. Yet players often associate communication within ad hoc teams with frustration and wariness. Though previous works have quantitatively analyzed communication patterns at scale, few have identified the motivations of how a player makes in-the-moment communication decisions. In this paper, we conducted an observation study with 22 \textit{League of Legends} players by interviewing them during Solo Ranked games on their use of four in-game communication media (chat, pings, emotes, votes). We performed thematic analysis to understand players' in-context assessment and perception of communication attempts. We demonstrate that players evaluate communication opportunities on proximate game states bound by player expectations and norms. Our findings illustrate players' tendency to view communication, regardless of its content, as a precursor to team breakdowns. We build upon these findings to motivate effective player-oriented communication design in online games.
\end{abstract}



\maketitle

\section{Introduction}

A team thrives and dies by its communication. Well-oiled communication is the engine that drives collaboration, underpinning crucial team processes. This holds particularly true for ad hoc teams, or swiftly formed temporary teams fulfilling a specific goal, as effective communication between unfamiliar members affects mission success~\cite{white2018, jarvenpaa1998, mesmer2009, marlow2018}. Thus, previous work indicates that ad hoc team structures may benefit from frequent, proactive communication to develop team cognition, enhance team cohesion, and strengthen interpersonal relationships~\cite{capiola2020, strater2008, roberts2014}. 

However, these communication strategies may not be applicable to ad hoc teams in virtual settings. Members in virtual ad hoc teams need to bridge their inherent physical and social distances without the aid of context-laden non-verbal cues~\cite{eisenberg2018, morrison2020}. Designing for optimal communication protocol in computer-mediated teamwork is not a universal experience. Conceptual models of ad hoc communication reveal that the specific characteristics of team context mediate positive outcomes of team communication~\cite{marlow2017}. For instance, increased communication openness and frequency can show a positive relationship with team performance~\cite{mesmer2009, monge2003}, but when the task centers around executing rapid, high-pressure decisions, explicit verbal communication may disrupt more vital team processes~\cite{entin1999, marlow2017}. Thus, each collaboration instance must factor in its virtuality and team characteristics to derive functional communication behaviors.

Online multiplayer games represent an especially challenging ad hoc collaboration domain for achieving effective communication. We situate our research on virtual ad hoc teamwork in \textit{League of Legends}\footnote{https://www.leagueoflegends.com/} (hereafter denoted as \textit{LoL}), a popular Multiplayer Online Battle Arena (MOBA) game that has been widely explored to understand team communication patterns in time-sensitive and high-intensity tasks~\cite{tan2022, zheng2023, leavitt2016, csengun2022players}. Previous work has identified communication antecedents of positive teamwork in MOBAs, such as communication sequences~\cite{tan2022} and hierarchical communication structures~\cite{kim2017}.

While previous literature has analyzed communication patterns through in-game data or isolated messages~\cite{tan2021less, tan2022}, communication processes in virtual ad hoc teams are more than just the sum of messages sent. There remains a gap in understanding what factors dynamically influence communication decisions in real time. Exploring the in-the-moment motivations behind communication decisions can reveal barriers to effective communication and illuminate how these decisions influence team collaboration in dynamic, real-world ad hoc contexts. Thus, this work seeks to capture the in-the-moment, player-centered perspective on how \textit{LoL} players decide to engage in different forms of communication within the game. We further connect these findings to evaluate how the communication experience and perception throughout the game affect the team members' attitudes towards their teammates, contributing to the ongoing discussions of communication and its effects on team trust and cohesion in virtual ad hoc teams. We thus approach this inquiry by first defining the forms and contexts of in-game communication. Then, we analyze the psychological and normative aspects that influence how players make communication decisions. We also investigate how communication processes shape players' perceptions of their teammates, focusing on the role these interactions play in fostering or hindering trust within the team.

In this paper, we address the following three research questions:
\begin{itemize}
    \item \textbf{RQ1. When and why do players engage in in-game team communication in \textit{League of Legends}?}
    \item \textbf{RQ2. How do players assess and react to in-game team communication in real time?}
    \item \textbf{RQ3. How does the player's experience with in-game communication processes shape their perception towards teammates?}
\end{itemize}

To this end, we conducted a qualitative analysis to identify context-dependent factors that drive communication decisions in real time. We performed an in-situ observational interview study with 22 players, observing their communication patterns and asking them to explain their communication choices during the game. This method directly engages with the player during their play, highlighting the proximate influences for their behavior. We follow in the footsteps of qualitative work by Buchan and Taylor~\cite{buchan2016} who looked at subjective player experiences in team coordination through interviews. However, we focus on the granular details of communication mechanisms and observe how they change and adapt throughout the game. We analyze players' usage and perception of the prominently used in-game communication methods, across both verbal (chat) and non-verbal (ping, emote, vote) methods. By incorporating both media, we paint a holistic picture of communication patterns and assess the natural trade-offs of each medium in different communicative contexts.

Our findings demonstrate that players evaluate immediately relevant in-game states, such as action cost and timeliness to make their communicative decisions, while bound by their expectations and norms set from repeated experiences. We show that players associate frequent communication with disruptive players, regardless of their communicative intent. Players instead virtue a teammate's commitment to the game, often demonstrated through action rather than explicit communication. Through this work, we motivate effective communication design that incorporates individual player context.

\section{Related Work}
Our work investigates in-game team communication in \textit{League of Legends} (\textit{LoL}), an online multiplayer game with a virtual ad hoc collaboration setting. Thus, we review prior work on communication in virtual ad hoc collaboration and communication patterns of online multiplayer games to situate our work.

\subsection{Communication in Virtual Ad Hoc Collaboration}
Team communication takes many different forms. Among them, a swiftly started collaboration between previously unknown members, referred to as an \textit{ad hoc} team, condenses communication actions into fast-paced, transient, and reactive teamwork~\cite{finholt1990}. Communication in ad hoc teams is critical to the mission outcome~\cite{white2018, jarvenpaa1998, mesmer2009}. Previous work has explored successful communicative practices adopted by unfamiliar members in various professional ad hoc disciplines, such as healthcare~\cite{roberts2014, chalupnik2020, evans2021}, military~\cite{pascual1999, capiola2020}, and software development~\cite{cherry2008}. Skilled ad hoc members working in high-intensity situations strategize their communication to establish rapport, coordinate tasks, and share critical information at opportune moments. These probes find that proactive and coherent face-to-face communication mediates swift trust among team members, which in turn influences team performance~\cite{capiola2020, strater2008}.

In comparison to traditional face-to-face contexts, virtual ad hoc teams are characterized by technological constraints that hinder effective communication. Computer-mediated communication in virtual, and often global, teams must overcome the physical, social, and cultural distances inherent to distributed work~\cite{oleary2007, morrison2020}. The limited communication channels of virtual teams put up significant barriers to cognitive alignment between strangers --- individuals working in temporary virtual teams experience reduced presence, individual identification, and immersion due to the lack of contextual cues~\cite{altschuller2013, altschuller2010}. Virtual collaborations without non-verbal signals (e.g., facial expressions and voice inflections) lead to weaker trust as members struggle to gauge intent, emotions, and engagement from their collaborators~\cite{morrison2020}. When members are unable to observe each others' actions, they may act based on their biased or inaccurate, negative perceptions~\cite{penarroja2013}. The absence of contextual information thereby implies that virtual ad hoc teams must frequently engage in explicit, task-oriented communication to facilitate coordination.

In reality, previous work reveals that the efficacy of virtual ad hoc team communication is shaped by the unique conditions of the collaborative context~\cite{marlow2017, mesmer2009}. Marlow et al.~\cite{marlow2017} identify three key communication constructs in their conceptual framework of virtual team communication processes: communication frequency, quality, and content. They emphasize the importance of analyzing these disparate constructs under the specific virtual context for a more granular understanding of effective communication processes. For example, while communication quality has often been linked to team cohesion, communication frequency shows varying relationships depending on the team composition and structure. Some studies find that overt sharing of information promotes positive team states~\cite{mesmer2009} and increased communication frequency is generally salient with team development and functioning~\cite{monge2003}. On the other hand, minimal communication can be more conducive to successful collaboration in team compositions where team members possess high levels of expertise and share a common understanding of the task at hand~\cite{entin1999}.

More specifically, communication in an online multiplayer gaming context parallels collaboration dynamics within virtual ad hoc teams working under significant pressure. Crisis management teams have been commonly used to explore team decision-making in dynamic and extreme situations~\cite{uitdewilligen2018crisis, altschuller2008potential}. These contexts include natural disasters~\cite{longstaff2008natural}, pandemics~\cite{white2020covid}, and critical operations like nuclear plant control~\cite{stachowski2009crises}. These extreme environments are high-pressure work settings defined by hostile conditions, isolation, limited time, and severe consequences for failure~\cite{harrison1984exotic, bell2016extreme}. Many of these characteristics are mirrored in online multiplayer games, where players face time limits and immediate, high-impact outcomes, which differ from traditionally studied, more stable contexts of ad hoc collaboration~\cite{musick2021cognition}. In such extreme contexts, Zijistra et al. revealed that communication patterns that are stable, balanced, and reciprocal lead to more effective collaboration~\cite{zijlstra2012interaction}. Similarly, Vinella et al. found through simulations of virtual ad hoc bomb diffusion teams that usage of role-aligned communication demonstrated higher team performance and perceived collaboration quality~\cite{vinella2022personality}.

Likewise, previous works have investigated communication patterns in various ad hoc teams. Although communications in \textit{LoL} may share similarities with previous work as the teams within can be viewed as virtual ad-hoc teams under high pressure, the unique contexts of \textit{LoL} may pose a different influence on the communication. 
Moreover, there still remains a gap in understanding how these communication patterns are driven by underlying individual cognitive processes in real time. Thus, understanding in-the-moment motivations behind communication decisions can provide insights into what deters teams from achieving effective communication and how communication influences team collaboration in real-world ad hoc contexts.

\subsection{Effective Communication Processes for Ad Hoc Teams in Online Multiplayer Games}

Gaming research has long aimed to map effective communication practices for ad hoc teams in online multiplayer games. Previous work has examined teamwork and communication in both cooperative games, such as puzzle platformers~\cite{tan2021less} and Massively Multiplayer Online Role Playing Games (MMORPG)~\cite{petter2011}, as well as competitive games, such as the First-Person Shooters (FPS)~\cite{tang2012verbal, taylor2012fps} and Real-Time Strategy (RTS) games~\cite{laato2024starcraft}. Communication processes in different game genres are defined and constrained by in-game affordances, gameplay mechanics and design, and play motivation. For example, MMORPG players' motivation centers on socialization and immersion, leading to more social-oriented interactions~\cite{bisberg2022contagion}. Competitive team games with short and intense game sessions emphasize goal-oriented communication: Tang et al. and Taylor observed reliance on callouts and codified speech for coordination among competitive FPS players~\cite{taylor2012fps, tang2012verbal}. In the asymmetrical horror game \textit{Dead by Daylight}\footnote{https://deadbydaylight.com/}, players formulate metacommunicative codes to overcome the lack of explicit communication channels; however, the ambiguity of interpretations caused frustration among players~\cite{deslauriers2020dbd}. This body of research highlights the diverse ways communication adapts to the unique demands and constraints of different game genres. 

Among these genres, Multiplayer Online Battle Arena (MOBA) games embody many components of challenging synchronous collaboration, including rapidly forming and dissolving team composition, high interdependence, and dynamic, high-stress environment. Various works have observed the prevalence of negative within-team communication in MOBAs, which undermines player performance and satisfaction ~\cite{monge2022effects, kou2014, canossa2021honor}. Additionally, non-verbal communication affordances (e.g., pings, emotes, animations) enable dispersion of communication across diverse channels according to situational needs. Despite --- or perhaps due to --- these complex factors, numerous works have examined effectual communication patterns in MOBAs. These works often use data-driven approaches to identify antecedents to successful or failed communication processes~\cite{tan2022, zheng2023, leavitt2016, csengun2022players}. For instance, Tan et al.~\cite{tan2022} have shown that positive chat sequences, such as apology to encouragement and suggestion to acknowledgment, improve team cohesion and teamwork in \textit{LoL}. The usage frequency of non-verbal pings in \textit{Heroes of the Storm}\footnote{https://heroesofthestorm.blizzard.com/} showed a concave relationship with player performance as it enabled swift and precise communication, but also became interruptions and distractions when too abundant~\cite{leavitt2016}. Meanwhile, Buchan and Taylor~\cite{buchan2016} qualitatively approached communication through the lens of the players' subjective experiences in MOBA games. They identified communication as a core category perceived to be influencing team play. The results showed that players associated excessive quantity of communication with negative team experiences, favoring no communication to excessive communication. These studies study the relationship between communication patterns and teamwork processes. But communication is not conducted in isolation --- communication processes are subject to the influences of the dynamic game environment.

Thus, our work builds upon the previous literature to identify in-the-moment communication decisions to uncover the individually motivated mechanics of player communication in MOBAs. We look at team communication in the context of \textit{LoL}, a well-established testing ground for probing team dynamics in temporary teams due to its clearly defined game parameters and rich in-game data~\cite{kou2014, kwak2015exploring, kwak2015linguistic}. By observing and inquiring about players' communication choices in \textit{LoL} during real-time play, we conceptualize the assessment processes that inform player communication behavior. We also offer a comparative analysis of when various communication media are used, offering new insights on relatively underexplored non-verbal communication modes such as emotes and votes.

\subsection{Communication Breakdown in Online Multiplayer Games}

Despite the efforts to foster effective collaboration and communication in online multiplayer games, in reality, cooperative online multiplayer games are plagued by team communication breakdowns. Though previously mentioned work has highlighted the pathways to achieve successful communication in online ad hoc teams, many obstacles block players from adopting such practices. Previous work has demonstrated that players desire and recognize productive communication behaviors within online multiplayer games~\cite{kou2014}. Yet, online multiplayer games, especially those of competitive genres like MOBA, are prone to within-team conflicts arising from unhealthy communication patterns between allies. MOBA games have frequently demonstrated interactions between players in which the communication becomes unconstructive, hostile, or abusive~\cite{kou2020toxic, beres2021, nexo2023players}. Aggressive and hostile communication patterns such as flaming and trolling often lead to a breakdown in team cooperation, causing emotional distress, threatening psychological safety, and decreasing overall team morale~\cite{kou2020toxic}.

To promote enjoyable gaming experiences and maintain player retention, game developers have made an effort to design supportive communication tools that expedite the information sharing process (e.g., pings and votes) or encourage social bonding and copresence (e.g., ``fist bump'' in \textit{LoL}~\cite{jarvis2024}). Yet, these tools take time to break into players' routines, and at worst, are misused for adverse behaviors. We also highlight the normative impact of gaming culture that blocks players' ability to engage in constructive communication. Previous work has shown that players exhibit dismissive behavior towards healthy communication, downplaying positive messages~\cite{poeller2023} and normalizing harmful communication~\cite{beres2021}. In this work, we explore the real-time and individual-specific challenges to realize effective communication during game sessions with unfamiliar teammates. Our work observes how communication, in conjunction with normative beliefs, informs the player's perception of their teammates.
\section{Study Context: League of Legends}
\textit{League of Legends} (\textit{LoL}) is a popular Multiplayer Online Battle Arena (MOBA) game, whose genre is defined by two competitive teams of human players battling for a common objective. In \textit{LoL}, two symmetrical teams of five members aim to destroy the other team's base (\textit{Nexus}). Each player selects a character from a pool of \textit{champions}, each equipped with unique abilities. Notably, ~\textit{LoL} game sessions are relatively short, generally lasting from 25 to 40 minutes. 

\begin{figure*}
    \centering
    \includegraphics[width=\textwidth]{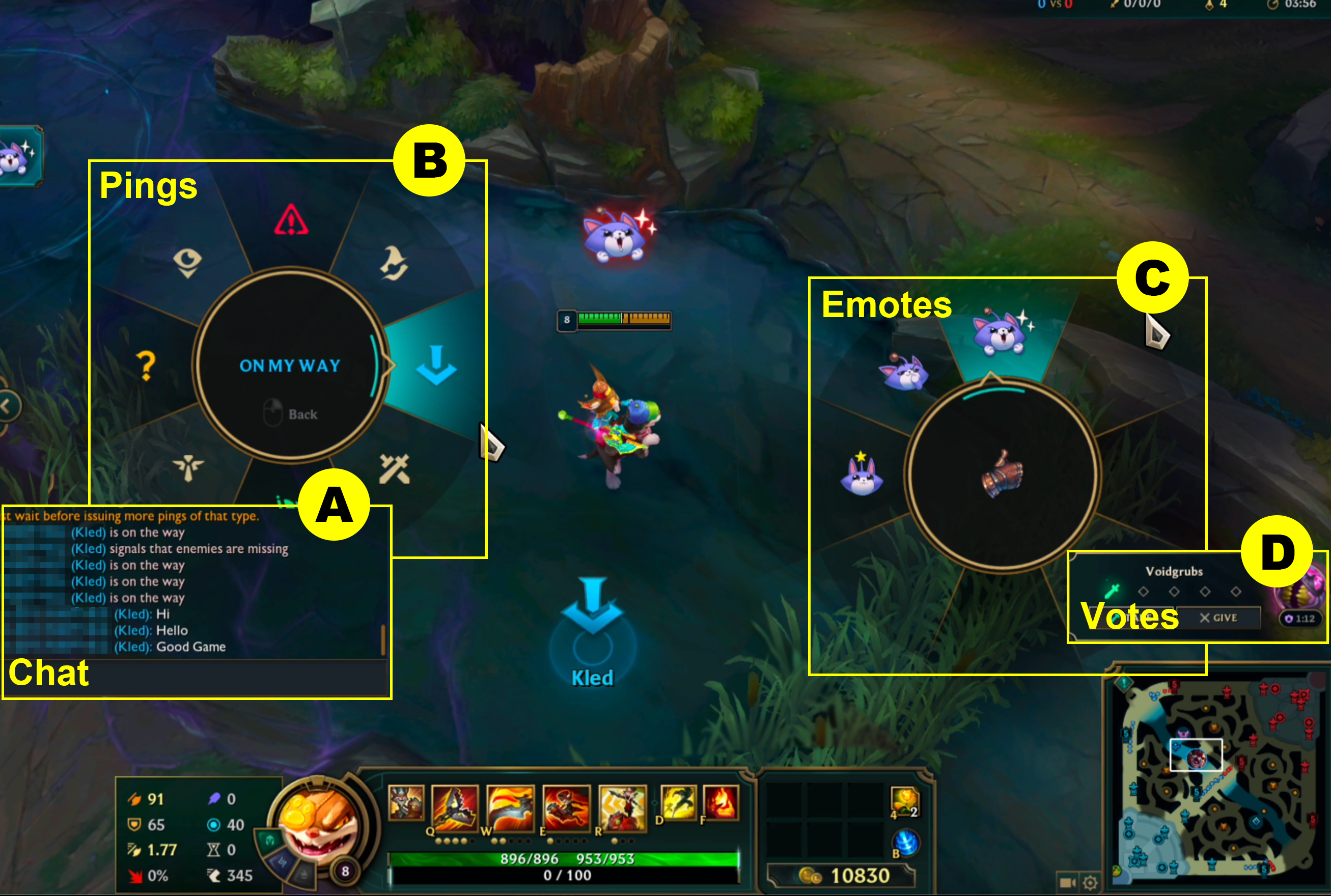}
    \caption{An example of four main communication modes of \textit{LoL}. (A) Chat is the medium through which players can type in-game messages to other players or read previous logs of in-game changes or signals, (B) Pings are quick alerts used for signaling information to other teammates, (C) Emotes are used to express emotions to other players, and (D) Votes are used to determine calls for objectives or to surrender the game.}
    \label{fig:modes}
    \Description{A screenshot of the gameplay screen of League of Legends, which demonstrates how chat, pings, emotes, and votes are shown in the game. The figure shows how each communication tool appears in the game, such as the chat box, ping and emote wheels, and the vote window on the right side of the screen.}
\end{figure*}

As \textit{LoL} is a \textit{competitive} team game, the outcome of the game depends heavily on cooperation between team members to achieve victory. The game supports real-time cooperation through diverse within-team communication channels native to the platform. In this paper, we investigate \textit{LoL} players' use of four main communication modes for corresponding with allies during the game: chat (verbal), pings, emotes, and votes (non-verbal). We illustrate how each communication mode may be used and represented in the game in Figure ~\ref{fig:modes}. The player base relies on these features to exchange key information, indicate intent, and express emotions throughout the entire session.

For verbal communication, players can type in the in-game chat before, during, and after the game (Figure ~\ref{fig:modes}A). Unlike other team-based competitive games such as \textit{Overwatch}\footnote{https://overwatch.blizzard.com/} and \textit{Valorant}\footnote{https://playvalorant.com/} or other MOBA games such as \textit{DoTA 2}\footnote{https://www.dota2.com/} and \textit{Heroes of the Storm} that offer voice chat for all teams, \textit{LoL} only enables it for a pre-formed party. Despite the potential benefits of voice-based communication for impromptu teams, \textit{LoL} developers have decided against voice chat, citing that it ``\textit{[does not solve] all behavioral issues and definitely introduces some new ones... Especially for women and POC (People of Color) who get unfairly targeted by simply participating in voice comms.}''~\cite{carver2023}

In \textit{LoL}, non-verbal communication is facilitated through pings, emotes, and votes. Pings are quick alerts used to signal information by placing markers on the map or characters. Players can access pings via a ping wheel (Figure ~\ref{fig:modes}B) or keyboard shortcuts. There are two types of pings: visual and UI pings. Visual pings, triggered by clicking the terrain or minimap, appear on the map and include generic markers for drawing attention and eight specific ``Smart'' pings (e.g., Retreat, On My Way, Assist Me) with predefined meanings shown in Figure ~\ref{fig:modes}B. UI pings share information about the status of the clicked interface elements, such as items or skills. Most pings, except non-targeted generic visual pings, are logged in the chat and accompanied by a distinct audio cue.

Emotes are expressive images or animations that convey emotions during the game, often featuring characters with various expressions like excitement, remorse, or provocations. When triggered, an emote appears above the player’s character for a few seconds and briefly in a bubble on allies’ screens. Unlike pings, emotes are visible to both allies and nearby enemies. Players can purchase emotes with in-game currency and customize their emote wheel (accessible via a shortcut) with up to nine options. Chat, pings, and emotes can be muted individually or entirely for specific players or everyone.

Finally, players can also communicate through surrender and objective votes. Starting at the 15-minute mark, a player may anonymously initiate a surrender vote, which appears on the right side of the screen. If at least 70\% of the team (or four players) vote ``Yes'' within 60 seconds, the game ends. If the vote fails, the team must wait three minutes to try again. In 2022, \textit{LoL} introduced objective voting, triggered when a player pings a field objective like Baron or Drake. This vote lets teammates decide whether to ``Take'' or ``Give'' the objective. We note that the chat system in \textit{LoL} textualizes all forms of communication. Chat messages, ping notifications, and in-game announcements are all interspersed in a single channel. 

These channels are used at different frequencies and at different points of the game. Ping usage is persistent and frequent throughout the game: the two most commonly used pings (On My Way and Enemy Missing) are used an average of $0.267$ and $0.164$ pings per minute respectively across all positions and ranks. There is a positive trend of increased ping with rank (reaching $0.489$ and $0.245$ pings per minute respectively for Master players and above)~\cite{log2024ping}, showing that as players become more skilled at the game, they are able to communicate more actively through non-verbal channels. In contrast, other non-verbal gestures are used sparingly. Though official statistics are not provided for votes and emotes, votes are limited by timers, which are tied to objectives (every 5–6 minutes), or surrender cooldowns (every 3 minutes). Emotes are used infrequently based on player norms and are often reserved for reacting to significant in-game events.

While previous works have often studied how players use one specific feature~\cite{tan2022, leavitt2016}, we aim to better understand how players use all of these features in conjunction. We illustrate the decisions players make in each communication attempt and how communication used by other team members shapes their team perception in real time.
\section{Methods}
This work uses qualitative methods to explore the communication behaviors of players in \textit{League of Legends (LoL)}. By qualitatively observing and inquiring about player communication decisions as they occur, we aim to extract insights into players' reasoning, strategies, and the underlying factors influencing their choices during the actual conditions of gameplay. We observe \textit{LoL} players during real ranked games, asking them in-the-moment questions as well as follow-up interview questions after the matches have ended to capture the nuances of their communication decisions. 

\subsection{Participants}
We recruited participants for the study through university forums and social media in South Korea. Participants were required to be 18 or older and active players of Solo Ranked mode in \textit{LoL} with a valid rank during the current season at the time of the experiment (Season 2024). The recruitment post notified participants of the observational nature of the study and informed them that they would be expected to speak out loud and answer questions during their play sessions. A total of 36 players completed the recruitment survey, which asked for a self-report of their age, game history, preferred roles, and current rank.

We conducted in-person interview studies with a sample of 22 players. This sample excluded players who had played for less than a year, who were not willing to answer questions during the game, or who were unable to participate in person. From the remaining pool, players were chosen to maximize the diversity and representativeness of the player base based on their rank, experience, and roles. If several players shared similar profiles, we randomly selected between the participants. We conducted 17 interviews through this sampling method. Out of the first 17 participants, 16 participants identified as male, and one identified as female. Consequently, to increase the gender diversity of the sample and ensure that the results reflect a broad range of player experiences and perspectives, we specifically recruited female \textit{LoL} players through snowball sampling, while maintaining diversity in preferred roles and rank. We recruited and interviewed participants from the survey responders until qualitative saturation was reached, following the definition by Braun and Clark~\cite{braun2021saturation}. The final sample consisted of 16 male ($72.7\%$) and 6 female ($27.3\%$) participants. This ratio approximates the imbalanced gender demographic of \textit{LoL}, where estimates have suggested that $80$-$90\%$ of the player base is male~\cite{kordyaka2023gender}. We address the influence of gender identity on communication in Section ~\ref{discussion} and ~\ref{limitations}. The full list of participants and their information is shown in Table ~\ref{table:participant_info}.

The players' age ranged from 20 to 32 years old (mean=$23.7$ years, SD=$3.3$ years) and players' \textit{LoL} experience ranged from 2 to 13 years (mean=$7.7$ years, SD=$3.9$ years). The Solo queue ranks of the players were 2 Iron ($9.1\%$), 2 Bronze ($9.1\%$), 5 Silver ($22.7\%$), 7 Gold ($31.8\%$), 5 Platinum ($22.7\%$), and 1 Emerald ($4.5\%$) at the time of the study. Though the distribution is not as even as the Solo queue rank distribution in the Korean \textit{LoL} server ($12\%$ Iron, $19\%$ Bronze, $16\%$ Silver, $15\%$ Gold, $18\%$ Platinum, $13\%$ Emerald, and $5\%$ Diamond and above~\cite{log2024}), it encompasses the diverse range of skills of most \textit{LoL} players. Thus, the selected players reflected a comprehensive sample of engaged players with varying experience and skill levels.

\begin{table*}
\centering
\caption{Participant Information and Game Session Information of \textit{League of Legends} Players}
    \label{table:participant_info}
\begin{tabular}{c|c|c|c|c|c|c} 
\toprule
\textbf{ID} & \textbf{Gender} & \textbf{Age} & \textbf{Experience} & \textbf{Solo Rank Tier} & \textbf{Role Played} & \textbf{Game Outcome}  \\ 
\hline
P1          & Male            & 24           & 12 years            & Silver                  & Jungle               & Win                    \\ 
\hline
P2          & Male            & 23           & 3 years             & Silver                  & Top                  & Loss                   \\ 
\hline
P3          & Male            & 32           & 12 years            & Bronze                  & Mid                  & Loss                   \\ 
\hline
P4          & Male            & 29           & 10 years            & Silver                  & Jungle               & Win                    \\ 
\hline
P5          & Male            & 27           & 11 years            & Emerald                 & Bot                  & Loss                   \\ 
\hline
P6          & Male            & 21           & 11 years            & Platinum                & Top                  & Win                    \\ 
\hline
P7          & Male            & 27           & 3 years             & Gold                    & Jungle               & Win                    \\ 
\hline
P8          & Male            & 20           & 9 years             & Gold                    & Bot                  & Win                    \\ 
\hline
P9          & Male            & 25           & 9 years             & Bronze                  & Jungle               & Loss                   \\ 
\hline
P10         & Male            & 23           & 7 years             & Silver                  & Jungle               & Loss                   \\ 
\hline
P11         & Male            & 21           & 12 years            & Gold                    & Mid                  & Loss                   \\ 
\hline
P12         & Male            & 22           & 11 years            & Platinum                & Mid                  & Loss                   \\ 
\hline
P13         & Female          & 24           & 3 years             & Gold                    & Support              & Loss                   \\ 
\hline
P14         & Male            & 25           & 10 years            & Platinum                & Bot                  & Win                    \\ 
\hline
P15         & Male            & 26           & 10 years            & Gold                    & Jungle               & Win                    \\ 
\hline
P16         & Male            & 26           & 13 years            & Gold                    & Mid                  & Loss                   \\ 
\hline
P17         & Male            & 25           & 8 years             & Platinum                & Support              & Loss                   \\
\hline
P18         & Female            & 21           & 3 years             & Platinum                & Support              & Loss                   \\
\hline
P19         & Female            & 20           & 6 years             & Gold                & Support              & Loss                   \\
\hline
P20         & Female            & 20           & 2 years             & Iron                & Top              & Loss                   \\
\hline
P21         & Female            & 20           & 3 years             & Iron                & Jungle              & Win                   \\
\hline
P22         & Female            & 21           & 2 years             & Silver                & Support              & Loss                   \\
\bottomrule
\end{tabular}
\end{table*}

\subsection{Procedure}
To capture the in-game mechanics of communication patterns and dynamically changing communication behavior in \textit{LoL}, we conducted an in-person observation and interview study. The study was conducted with the approval of the Institutional Review Board at the first author's research institution.

\subsubsection{Study Environment}

Each participant was asked to play a Solo Ranked game of \textit{LoL} while researchers observed and inquired about their actions in real time. In the process of study design, alternative study setups were considered. An initial plan to observe participants remotely through screen sharing was discarded as pilot studies revealed that latency and network issues were significantly disruptive to the researchers’ ability to observe and ask questions in a timely manner. The research team also decided not to recruit participants to observe in their own homes due to concerns about privacy and safety. Thus, to better gather responses to in-the-moment inquiries from the participants without any latency, the study was held in person within a controlled research environment to enhance the clarity and quality of the question-and-answer process. This approach also allowed researchers to note the participant's gaze, hesitation, and other subtle movements that would be missed in remote settings. Though an in-lab study does not perfectly recreate the in-home environment, the research team determined that this option best balanced the various tradeoffs.

The research team worked to ensure that the study environment best suited each participant’s preferences in the following steps. We first conducted several pilot interviews to tailor the play space to prioritize player comfort and approximate real-life conditions. The study took place in an enclosed room --- frequently used for interviews and user studies --- equipped with large desks and office chairs. We limited natural light and turned on overhead lights for visibility. Moreover, players were individually asked if the environment felt natural and comfortable, and adjustments were made based on their feedback. We set up the study environment with equipment (mouse, keyboard, monitor) designed specifically for online gaming. The players were also permitted to bring personal equipment if they wished. Before entering the game, the players were instructed to adjust both the equipment (e.g., mouse sensitivity) and in-game settings (e.g., shortcut keys). All players were given as much time as needed until they expressed satisfaction with the setup. At the end of the study, we asked players if anything about the setup or procedure had negatively impacted their gameplay. Three players reported feeling some discomfort from using unfamiliar equipment but noted that it did not affect their typical playstyle or their answers. The participants were compensated 20,000 KRW (approximately 15 USD) for completing the study.

\subsubsection{Interview Process}

The researchers observed the game session through a separate screen connected to the player's monitor and noted any communication actions, attempts, and responses by the player. During the study orientation, researchers emphasized that participants should play and communicate naturally, including using offensive language, muting or reporting other players, or forfeiting the game if desired. Participants were assured that all data would be anonymized for analysis. To minimize distractions, they were informed that they could skip questions if they found them intrusive or preferred not to respond. During intense in-game situations in which the participant could not answer, the researchers documented the context and either repeated the question once the game state had stabilized or immediately after the match ended. We discuss the limitations of using an observational approach to study in-game communication, such as social desirability bias~\cite{grimm2010social}, in Section ~\ref{limitations}.

The researcher observed the communication between teammates, noting what triggered the communication or what communication medium was used for different purposes. Based on these observations, the participant was asked questions about why they did or did not perform certain communication actions to understand their assessment and perception of the communication. Some of the questions were asked to all participants, such as the reasoning behind the frequency of using certain communication media and their perception towards teammates who engage in certain forms of communication. Other non-structured questions were asked when the player triggered a certain action (``\textit{You just pinged your ally with Enemy Missing ping multiple times. What was the purpose?''}) or responded to (or ignored) their team's communication (``\textit{It seems that you opted to not vote for the surrender vote. Why is this so?}'').

The researchers recorded the gameplay and thoroughly transcribed observations and game states during the game. The observations included types of communication media used (chat, ping, votes, emotes), the target of the communication (if unclear, then players were asked to clarify the target), player reactions to ongoing discourse or communication usage by their team, and physical reactions such as spoken utterances and body gestures. After completing the game, participants engaged in a 15 to 20-minute post-game interview. They were asked to reflect on their in-game communication behavior and perceptions, including the motivations behind their communication tendencies and choices. The interview also inquired how their teammates' communication behaviors influenced their perception of those teammates, as well as the overall impact of such interactions on their mental state or performance. Participants were further invited to share suggestions for improving communication in \textit{LoL}, such as the potential addition of voice chat. The full list of the interview questions is provided in Appendix ~\ref{appendix2}.

Overall, a total of 24 games were played, of which two were forfeited within 20 minutes. Players were asked to play another game if their first game ended within 20 minutes due to a surrender vote from either team as these games did not demonstrate communication across all stages of the game. The other 22 games lasted a minimum of 24 minutes, most of which were played to completion without forfeit from either team or with a forfeit when the victor was very clearly determined.

\subsection{Thematic Analysis}
We conducted an inductive thematic analysis applying the methodology from Elo and  Kyngäs~\cite{elo2008qualitative}. The researchers gathered the transcript of the in-game and post-interview, video recording and the replay file of the game, and observational notes for each participant. We incorporated the notes on participant behavior, game states, and other players' communication patterns into the transcript at its corresponding times, providing contextual information on what was happening at the time of the question or the player's reaction.

Before initiating coding, the first, second, and third authors familiarized themselves with the data collected. The three authors then independently performed line-by-line open coding on eight participants' data to identify preliminary themes. After the initial coding was completed, the authors shared the codes to combine convergent ideas and discuss any differing perspectives. The first author then validated the codes on the remaining data, engaging in discussions with the second and third authors to iterate on the codebook. The final codebook contained 55 codes with 13 categories, organized by themes that answer each research question in depth. For RQ1, we find themes of communication types and what triggers or deters a player's decision to communicate. For RQ2, we categorize factors used by players to assess communication opportunities as well as reactions based on such assessment. For RQ3, we find themes on how the team's communication behaviors affect a player's perception towards other teammates during the game. We provide the final codebook in Appendix ~\ref{appendix}. 
\section{Results}
We first describe communication patterns within the full chronological context of the game in \textit{League of Legends (LoL)}, separated into four sections based on changing coordination dynamics. Based on this context, we identify core factors players assess to decide when to participate in communication with other teammates. Afterward, we discuss how communication shapes player perceptions toward their teammates, showing player's wariness towards players actively engaging in communication. 

\subsection{Communication Patterns in Context}

We discuss the communication patterns among teammates within the game. We organize the data into chronological phases of the game for a structured analysis of how the context shapes communication patterns. 

\subsubsection{Pre-game stage}
Before gameplay begins, team communication opens with \textit{team drafting}, where players are assigned roles (Top, Mid, Bot, Support, or Jungle) and take turns picking or banning champions. In Solo Ranked mode, roles are pre-assigned based on player preferences selected before queueing. Once teams are set, all players enter \textit{champion select} stage, alternating champion picks and banning up to five champions per team. During this stage, communication is limited to text chat. The usernames are anonymized (i.e., replacing the name with aliases) to prevent queue dodging by checking third-party stats sites such as OP.GG\footnote{https://www.op.gg/}, leaving the chat as the only option to inform individual strengths and preferences. 

Team composition in \textit{LoL} is crucial to the strategy and outcome of the game~\cite{ong2015player}, setting the basis for future interactions. Most participants acknowledged the importance of balanced and synergistic team composition, especially as players move into higher ranks where team coordination outweighs individual excellence. Yet, we observed a distinct lack of verbal communication between the members during this period across all ranks. Participants attributed the lack of willingness to initiate a conversation on the dangers of starting the game on a bad footing. They prioritized ``not creating friction'' during this stage as negative impressions can propagate throughout the game. Some participants attempted communication to reduce such friction, such as P14, who stated,``\textit{If I had the time, I wanted to say that I will be banning [this Champion], just in case a player on my team wanted to play them.}'' However, several participants viewed any communication during the pre-game phase with wariness, as dissatisfaction or conflict at this step portended negative interactions between players in the game (P3, P9, P15). Thus, even when participants expressed doubt about other teammates' unconventional or non-meta champion picks, they refrained from entering into discourse. This contrasts with findings by Kou and Gui~\cite{kou2014}, which showed players attempt to maintain a harmonious and constructive atmosphere through greetings and introductions.

Another emergent code of the reason for not engaging in communication in the pre-game stage stems from different purposes of playing the game (P1, P5, P13, P16, P17). Despite being in ranked mode, which is more prone to increased competitiveness and effort, participants showed differing goals and levels of interest in winning the game. Several players stated that they had previously exerted great mental load in coordinating synergistic plays, but stopped as they gave less importance to winning at all costs (``\textit{I don't really play to win. I play \textit{LoL} to relieve stress, so I don't engage in chat.}'', P5). These players saw verbal communication with the goal of coordination as an unnecessary or even cumbersome component of the pre-game stage.

\subsubsection{Structured phase}
In many MOBAs, including \textit{LoL}, the early stages of the game play out in a formulaic manner: players join their lanes (Top, Mid, and Bot/Support), defeat minions to gain gold, buy items towards certain ``builds'', kill or assist in early objectives (Jungle), and battle counterparts in their respective lanes. Participants at this stage expressed that most players possessed tacit knowledge of what must be done, such as knowing when to aid their Jungle to capture a jungle monster, choosing the opportune moments to leave their lanes, or positioning wards (i.e., a deployable unit which provides a vision of the surrounding area) at the ideal placements. The participants assumed each player knew their ``role'' to fulfill, often comparing it to ``doing their share'' (P1, P3, P7, P19). In line with this belief, players rarely initiated preemptive or proactive verbal communication for strategic or social purposes at the early stage. 

Pings, on the other hand, constantly permeated the game. At this stage, players used ping to provide information relevant to others from their position, such as letting others know if an enemy went missing from their lane. As the players are largely separated and independent from one another, pings (coupled with the minimap and scoreboard) served as the primary channel for maintaining context over the game rather than as warnings or direct guidance to the players. For other non-verbal gestures, while objective votes would occasionally appear, they were rarely answered. Instead, relevant players near the objective would place pings or move toward it to help out their teammates.

Participants viewed the structured phase as a routine, but uncertain period of the game where the pendulum could swing in either team's favor. Players --- especially Jungles who roam the board looking for opportunities to ambush the enemy team in lanes (``gank'') --- sometimes felt hesitant to make calls and demands at this stage since ``\textit{[they] could make a call, but if I fail, they'll start blaming my decisions down the line.}'' (P7) But at this stage, participants believed that they held personal agency over the final game outcome. P1 and P6 stated that they entered the game with the mindset that only they had to succeed regardless of the performance of their teammates. This belief was reflected in their chatting behavior, where players prioritized focusing on their circumstances over the team's (``\textit{I mute the chat so that I don't get swayed by the team, as I can win the game if I do well.}'', P9).

\subsubsection{Group engagement phase}
As the game enters its middle phase, it provides opportunities for more diverse decision-making. Players may swap lanes, seize or trade crucial objectives, and fight in large battles involving multiple champions. At this point, teams typically have a clear outlook on which players and team have the advantage, requiring more team-driven decisions to maintain or overcome their current standing. Thus, players used verbal communication to discuss more complicated tactics that could not be effectively conveyed through pings.

But more often than not, chat messages became judgment-based. As enemy engagement with larger groups occurred more frequently, the availability for chatting would come after death, which led to comments on past actions rather than future choices. Additionally, the respawn timer for deaths becomes longer as the game progresses, providing more time to observe other players than in earlier phases. This gave players more opportunities to express dissatisfaction specifically towards certain plays, such as placing Enemy Missing pings on the map where other teammates are located to bring attention to their questionable play.

This stage also gave much more exposure of each other to the allies as the team would gather at a single point, giving way to greater scrutiny by their teammates. Repeated or critical mistakes put participants on edge, as they braced for criticism from their teammates. They expressed relief or surprise when the chat remained silent or civil, with P8 stating ``\textit{I messed up there. No one is saying anything, thankfully.}''

\subsubsection{Point of no return}
Meanwhile, verbal communication flowed out when the game had a clear trajectory to the end. Previous research has shown that both toxic and non-toxic communication skyrockets near the end of the game~\cite{kwak2015linguistic} when the players have determined the game outcome with certainty. We saw that this phase opened up both positive and negative sides of communication for guaranteed win and loss, respectively. The winning team would compliment and cheer each other through chat messages and emotes, while the losing side devolved into arguments and calling out. The communication at this stage was driven by emotion, showing excitement or venting frustration.

\subsection{Communication Assessment Process}

We describe the factors that users mainly focus on to assess when or when not to involve themselves in communication with their teammates. 

\subsubsection{Calculating communication cost}
One of the most proximate factors behind when communication is performed is the limited action economy of the game. In \textit{LoL} and other MOBAs, players can rarely afford time to type out messages due to the fast-paced nature of the game. In time-sensitive scenarios, the time pressure makes communication particularly costly. It is therefore unsurprising that much of the communication occurs after major events (e.g., battles and objective hunting), as players are given more downtime while waiting for teammates or enemies to respawn or regroup.

For periods where players were still actively involved in gameplay, the players made conscious decisions on choosing which communication media to use based on the perceived action availability and the importance of communicating the message. Players relied on pings for non-critical indications, believing that the mutual understanding of the game would get the message across. However, many players recognized that pings were prone to be missed, ignored, or misinterpreted by their allies (P2, P9, P16, P17, P20). Subsequently, participants typed out information considered to be too important to the situation to be misunderstood or missed by other players even if it caused delays in their gameplay (P10, P11, P14). Simultaneously, the priority of importance constantly shifted --- we observed multiple times participants start to type, but stop to react to an ongoing play, only to never send out their message.

\subsubsection{Evaluating relevance and responsiveness}
When the brief window of communication opportunity is missed, players are unlikely to ever send out that information. In \textit{LoL}, situations can change within seconds and certain communication media cannot keep up with the changing state of the game. For example, almost all study participants did not participate in votes for objectives. Among the tens of objective votes initiated among all the games in this study, no objective vote saw more than three votes, frequently being left with no vote beyond the player who initiated the vote. Some players, when asked why they did not participate, stated that the votes they made often became irrelevant as the game state had changed during the time it took to vote (P2, P11). Other players also discussed how information conveyed through communication can get outdated fast (P1, P8, P9). 

\begin{quote}
I can't always follow through with what I say [in the chat] since the game is really dynamic. My teammates don't understand such situations, so I tend to not chat proactively. - P9
\end{quote}

Thus, some players instead preferred to react through direct action (P8, P10, P11, P16, P20). P10 stated, ``\textit{I think it's enough to show through action rather than [using objective voting]. I can look out for how the player reacts when I request something from them.}''

On the other hand, such action-based responses left the player to assess whether and how the communication was received. P10 stated that they tried to predict whether a player understood their ping direction by how they moved, but it was hard to interpret their intent: ``\textit{members sometimes seem to move towards me but then turn around, and sometimes they even ping back but don't come.}''. P16 discussed how they weren't sure whether the ping was received, but performed it anyway since it felt helpful.

Similarly, participation in surrender votes (or lack thereof) carried different intent by the player. During most of the games that ended in a loss, one or more surrender votes were called by the participant's team. However, only two surrender votes achieved four or more players' participation. However, the reasons why a player chose to not participate varied. Some had decided to wait and see how other teammates voted, which may have paradoxically led many members to not participate in the vote (P4, P9). Meanwhile, others didn't reply as they didn't think the vote was actually calling for a response: P13 stated, ``\textit{I didn't vote because they were just showing their anger. It's just a member venting through a surrender vote that they're not doing well.}''

\subsubsection{Balancing information access and psychological safety}
While recognizing that communication would be useful or even necessary in certain situations, participants also put their psychological safety first over information access. Some players, worn down by the normalization of toxic communication such as flaming, muted the chat (P1, P9).

Many participants expressed the sentiment of ``protecting [their] mentality'', describing how certain communication harmed their psychological well-being. This communication did not always refer to negative communication; P9 often muted players who gave commands as they did not want to be ``swept up'' by others' play-related judgments. This separation even extended to other more widely considered essential communication forms, such as pings. Even after acknowledging that pings were vital and useful to the game, P9 went as far as muting the ping of the support player in the same lane after they sent a barrage of Enemy Missing pings that signified aggression and criticism. 

Additionally, the abundance and high frequency of communication also strained the limited mental capacity of the players. Many players, when asked why they had not replied to an objective vote or other chat messages, stated that they simply did not notice them among other events happening (P1, P2, P3, P9, P12, P15, P18, P20, P21). The information overload caused stress and became distracting to players.

\subsubsection{Reducing potential friction}
As demonstrated in the pre-game stage of the game, players sometimes used communication to minimize friction between their teammates. Some participants sacrificed time to apologize to other players when they believed themselves to be at fault. When asked why, P12 replied, ``\textit{There are too many people who don't come to help gank if I don't apologize.}''. Similarly, P5 sacrificed time typing in an apology after a teammate had died despite still being in the middle of a fight as they didn't wish to give the other player a reason to start an attack.

However, some noted that silence is sometimes the best answer to a negative situation. P4, after dying to the enemy, put into chat ``Fighting!'' (roughly meaning, ``We can do it!''). They stated ``\textit{I don't know why I do it... it probably angers [my teammates] more.
}'' They also stated that ``\textit{for certain people, talking in the chat only spurs them more. You just have to let them be.}'' Other players shared similar sentiments that being quiet and dedicating focus to the game was a better choice (P1, P11, P14).

For female players, the fear of gender-based harassment shaped their communication patterns. While \textit{LoL} does not provide any demographic information of a player to other players, almost all female participants noted experiences of receiving derogatory remarks or doubts about their abilities based on other players' assumptions of their gender, a trend frequently seen in male-dominated online gaming cultures~\cite{fox2016women, norris2004gender, mclean2019female}. They noted that the players were able to correctly guess their gender when the participant's role and champion fit into the preconceived notions of what women ``tended to play'' (i.e., female-identifying support champions, such as Lulu) or their username ``seemed feminine'' (P18, P19, P20, P21, P22). This led to certain players adopting tactics that signaled male-like behavior, such as changing their speech style to be more gender-neutral or male-like (P19, P21) and changing their username to sound more gender-neutral. Cote describes similar instances of ``camouflaging gender'' as one of five main strategies for women coping with harassment~\cite{cote2017coping}. However, some players opted to keep playing their preferred character or maintaining their username even if it signaled their gender, such as P21 who expressed, ``\textit{I cherish and feel attached to my username, so I don’t want to change it just because of [harassment and inappropriate comments].}'' These players valued self-expression and identity even at the risk of increased risk to unpleasant communication experiences.

\subsubsection{Forming performance-based hierachy}
Naturally formed leadership has often been observed in other works on \textit{LoL} teams~\cite{kou2014}. Kim et al. showed that more hierarchy in in-game decision-making led to higher collective intelligence~\cite{kim2017}. While they used ``hierarchy'' to mean varying amounts of communication throughout the game, we observed that the hierarchy extends further to performance-based hierarchy, where teammates in more advantageous positions are given greater weight when communicating with other players. Players actively chose to refrain from suggesting strategic plans when they were ``holding down the team'', recognizing that they held less power and trust among the team members (P8, P10, P12, P14, P22). The player who was losing against the enemy team was viewed as having no ``right'' to lead the team, which was reserved for well-performing players.

\subsubsection{Enforcing norms and habits}
One of the most common answers to why players performed certain communication actions, especially non-verbal actions such as pings and emotes, was ``a force of habit'' (P6, P7, P8, P9, P10, P12, P17). Players formed learned practices of using communication channels at certain points by observing other players exhibit the same behaviors. This promoted, for example, replying to an emote sent by the teammate with their own or pinging readied skills and items to emphasize relevant information for other players throughout the game. 

On the other hand, this meant that players were averse to communication patterns outside of the norm --- participants stated that they had a hard time adapting to new forms of communication, seeing no immediate benefit or impact from using them (P1, P8, P14, P13, P15, P17). Most egregiously, the recently introduced objective pings were largely viewed to be awkward to use and unnecessary (P1, P4, P8, P12).

\subsection{Impact of Communication Assessment}
We describe how the communication patterns and assessment of the players impact the individual players' perspectives on team dynamics.

\subsubsection{Relationship between trust and communication frequency}
Most participants saw value in constant and well-informed communication but with an important distinction: verbal communication with strangers rarely ended well. Players largely recognized frequent verbal communication to burgeon conflict, regardless of the message within. Even when players understood the helpful intent behind positive messages from the players, they compared actively talking players to be possible bad actors who were likely to exhibit toxic behaviors when the game turned against them. (P1, P4, P8, P12, P14)

\begin{quote}
I need to make sure to not disturb Twisted Fate. I saw him start to flame. It's not because I don't want to hear more criticism. I know these types. The more I react and chat with them, the more deviant they will become. - P4  
\end{quote}

Similarly, P19 lamented that players used to socialize more in the chat during the pre-game phase to build a fun and prosocial environment, noting a memorable example of encouraging each other to do well on their academic exams, but noted that such prosocial behavior has become much rarer during the recent seasons. They noted that there are inevitably players ``who take it negatively'' and thus stopped proactively typing non-game related messages in the chat.

Ultimately, players desired assurance and trust of player commitment. The participants trusted actions more than words to prove that the player remained dedicated to the game. Both P10 and P17 pointed out that it was easy to tell who was still ``in the game'' and motivated to try their best and that ``staying on the keyboard'' likely meant that they weren't invested or focused on the game. Players viewed such commitment to be the most important aspect of a ``good'' teammate, sometimes even more than their skill or performance (P9, P14). It is interesting to note that unlike what previous literature may suggest~\cite{marlow2018}, players' averseness to talkative teammates had less to do with the cognitive overload or distraction caused by the frequent communication, but rather due to the threats of future team breakdown. This view in turn also affected how players decided to communicate or not, as they believed that players would not take their suggestions or comments in a positive light.

\subsubsection{Perception of player commitment and fortitude}

Communication also acted as a mirror of their teammates' mental fortitude. A number of players mentioned how they valued a resilient mindset in their teammates playing the game, referring to players who remained committed to the game until the very end. They saw players who provoked or complained to teammates as ``having a weak mentality'' who had been altered by the bad outcomes of the game to act in an unhelpful manner towards the team through their communication. The communication actions of the teammate informed the participants of how steadfast their teammate remained in disadvantageous situations.  

\begin{quote}
It's not like I constantly reply in the chat or anything, but I pay attention [to the chat] to grasp the overall atmosphere of the team. If the team doesn't collaborate well then we lose, so I try to have a rough understanding of the mentality of the other players. - P13
\end{quote}

There were also instances of communication that helped players maintain a positive view of their teammates. For example, P11 mentioned near the beginning of the game, ``\textit{Looking at the chat, Varus player has strong mentality [for being so positive]. There were lots of points [in his support's] plays that he could have criticized.}'' Unfortunately, this view quickly soured when the Varus player devolved into criticism later in the late game phase where the Varus player started criticizing the support and other players. P11 then noted that the Varus player seemed to merely be ``bearing through the game''.
\section{Discussion} \label{discussion}
The results identify core factors of communication assessment and the effect of such assessment on the player's perception of the team. First, we discuss how our results provide deeper insight into the connection between trust and communication in ad hoc team communication. We also compare findings on communication in \textit{League of Legends (LoL)} that may be applied to other virtual ad hoc contexts. Then, we discuss how player values and priorities shape player engagement and assessment of communication. Finally, we offer insights on analyzing team processes through ad hoc teams in gaming environments.

\subsection{Relationship Between Trust and Communication Processes}

Many disciplines have aimed to analyze how communication structures affect team performance~\cite{chalupnik2020, finholt1990, roberts2014, capiola2020}. These works provide communication characteristics and signals that lead to more effective collaboration. However, realizing effective communication processes in swiftly starting teams is difficult, when team and individual cognitive processes may be particularly sensitive to the extreme conditions of the task environment. We thus aim to understand \textit{when} a user uses certain communication features, and more importantly, \textit{how} they make such a choice in the moment. We build upon works that have looked at verbal~\cite{tan2022, tan2021less} and non-verbal gestures (e.g., pings)~\cite{leavitt2016} in games as an indicator of team metrics. In turn, we offer a holistic view of players' communicative decision-making process, providing insight into how communication takes form in real time for a resource-limited, high-stakes environment.

In doing so, we uncover the role of trust driving the cognitive processes behind individual communication decisions. Our results find that the presence of communication signals to players the possibility of future team breakdowns. Previous works have also found that players prefer silence over excessive or negative verbal communication~\cite{buchan2016, tan2021less}. However, our results show some players showed an aversion to communication regardless of its content, even if the content is positive, wary of the communicator becoming hostile as the game progressed. Their behaviors and attitudes toward communication suggest that players' evaluation of harm from communication outweighs the possible benefits of positive interactions. Players of MOBAs, and online multiplayer games in general, join thousands of teams across their playtime, each composed of different members. This opens up the possibility for players to harmful communication at each match, and as players' past experiences with the negative impact of communication are put in the forefront of their minds, this may cultivate the reluctance to engage in communication of any kind.

We interpret these results under the framework of team processes-trust relationship. Research has demonstrated relationships between trust and team performance~\cite{erdem2003cognitive, kanawattanachai2002dynamic}, particularly in virtual teams~\cite{powell2006antecedents}. Mayer et al. defined trust as the willingness of one party to place themselves in a position of vulnerability to another, based on the expectation that the other party will carry out actions deemed important to the trustor, even without the ability to monitor or control their behavior~\cite{mayer1995trust}. According to this definition, we find that players used communication patterns as a gauge of trust in teammates to maintain commitment to the game. Wilderman et al.'s multilevel framework of trust encompasses multidimensionality and fluidity of trust~\cite{dirks2022trust} by breaking down the steps of how trust and team processes shape one another based on input-mediator-output-input model~\cite{wildman2012trust, lines2022meta}. The model posits that initial surface-level and dispositional inputs shape cognitive, affective, and attitudinal (trust) mediators, which influence team processes and performance, creating a feedback loop that adjusts trust in the team over time. From the perspective of this framework, player's previous negative experiences with team breakdowns (input of ``imported information'') and teammates' behavior alignment with these experiences (input of ``surface-level cues'') shape their perception towards the teammate (mediator of ``trust-related schema''), which in turn will shape the team process (output).

This result underscores the critical role of communication processes in fostering trust and highlights the need to examine players' in-the-moment motivations to understand how trust dynamically shapes and is shaped by team processes. By focusing on players' real-time decisions and actions, we can uncover how immediate factors --- such as perceived alignment with team goals, reactions to setbacks, or spontaneous communication efforts --- influence the development or erosion of trust. Such insights are essential for designing interventions and systems that promote adaptive teamwork and sustained collaboration in high-pressure, fast-paced environments like online multiplayer games.

\subsection{Defining Communication Processes Beyond Gaming Contexts}

Numerous multiplayer online games have been analyzed to study team communication, spanning First-Person Shooters (FPS)~\cite{tang2012verbal, taylor2012fps}, Real-Time Strategy games (RTS)~\cite{laato2024starcraft}, Massive Multiplayer Online Role-Playing Games (MMORPG)~\cite{petter2011}, and more. These games have different team structures (competitive vs cooperative), conflict types (Player-vs-Player or Player-vs-Environment), team persistence (short vs long-term), and communication channels (e.g., in-game and third-party voice chat, text chat, pre-defined messages, pings, reactions). We explain the distinctions of \textit{LoL} communication processes shaped by these differences and extend the findings to other gaming and virtual ad hoc domains.

As a competitive game, \textit{LoL} players place great importance on understanding how players are doing in comparison to the enemy team. A core design philosophy of many MOBAs is that players can easily quantify the game state. Kou et al. point to how the quantified views of the player and game state (such as rank and average kill rates) can lead to increased player stress and tension within the team~\cite{kou2018self}. Though Kou limits quantification as information gathered using third-party tools, we observe how the importance and availability of information access may shape communicative processes. In FPS games like \textit{Halo 3}, players rely on callouts from teammates to make crucial plays --- thus, disruptions to receiving these messages or incorrect calls can be a significant detriment~\cite{tang2012verbal, taylor2012fps}. On the other hand, \textit{LoL} players prioritize maintaining team integrity even at the expense of losing information, as reduced player commitment causes far more significant consequences for the game.

These differences in team contexts should be considered when designing and implementing communication features. In professional domains such as healthcare~\cite{roberts2014, chalupnik2020, evans2021} and military~\cite{pascual1999, capiola2020} where the consequences of every action may be critical, two-way communication is advantageous~\cite{zijlstra2012interaction}. However, in online gaming, layers often find communication to fall behind the quickly changing state of the game or remain unanswered. Thus, rather than communication that fades away, fast-paced games may want to adopt ``statuses'' that are persistent throughout the game. Finally, instead of sharing all information with every individual, information that caters to the player may reduce mental load and bring attention to critical information. Meanwhile, new forms of communication should be not easily abused to threaten the psychological safety of the players through misuse, such as \textit{LoL}'s bait ping~\cite{ramun2023} that was introduced and removed due to overtly toxic use of the new feature. 

\subsection{Understanding Player Values, Priorities, and Motivations in Team Communication}

On the surface, it may seem like all players in \textit{LoL} are working towards the same primary goal: winning the game. However, even in Solo Ranked mode where players are more extrinsically motivated by a performance-based ranking system, we find that players hold different, individual-driven goals and values, which can affect their in-game behavior and perceived experience~\cite{bruhlmann2020motivational}.

The findings on the communication assessment process show that participants prioritize different team and individual processes of the game. Some players value their psychological safety, forgoing access to game information in the process. Others instead approach the game with the drive to win, even if it means spending more time communicating important information or putting out possible firestarters of conflict.

Female players face unique challenges in their communication processes in \textit{LoL}. Female players must adopt coping strategies to prevent gender-based harassment from other players. In line with previous work, female participants described previous experiences of sexual harassment, flaming, and other verbal abuse regarding their gender identity~\cite{fox2016women, norris2004gender, mclean2019female}. To protect their psychological well-being, the female players took active steps to conceal their gender identity, such as changing their communication style and username to sound more gender-neutral. However, we also highlight that several female players prioritize and value the expression of their identity and preference during team processes. Safety may be an important aspect of communication, but communication process designs should also consider that players' priorities and values may not align with the most ideal or effective forms of communication.

Furthermore, the communication behavior of the players may reflect their unique cultural influence. \textit{LoL} is a popular and widely spread game for many Korean players~\cite{turbosmurfs}. In South Korea, playing \textit{LoL} is often thought of as mainstream culture~\cite{kukinews2019}. Nearly 70\% of all Korean male adolescents play \textit{LoL}~\cite{newsis2024}. For Korean players, \textit{LoL} represents a virtual social platform that enables the strengthening of social bonds between friends. However, in turn, the presence of such social connections may lead to increased pressure to perform better, as their in-game performance may influence a player's real-life identity. Such social bonds thus may exacerbate the competitiveness of players and perpetuate and normalize certain communicative behaviors. Another unique aspect of South Korean \textit{LoL} culture is the prevalent use of internet cafes, locally known as \textit{PCBang} (meaning ``Personal Computer Room''). These are spaces where players pay money by the hour to play games using high-quality computer equipment. This culture may also increase the need for players to end the game quickly and ``give up'' more easily, as longer games tie directly in with monetary loss.

\subsection{Merits and Pitfalls of Analyzing Teamwork through Online Multiplayer Games}

Online multiplayer games have been a ripe ground for understanding team dynamics, specifically for virtual, temporary teams~\cite{kou2014, kwak2015exploring, tan2022}. Clearly defined goals and boundaries of online games make it an ideal space to study interpersonal interactions and relationships for collaboration. The rich data from in-game statistics especially open it up for data-driven approaches that measure the impact of communication behavior on the team outcome. Analyzing teamwork in the online multiplayer gaming space enables easier and more direct control of the conditions present in team characteristics. It also allows analysis of teamwork across different skill levels and virtuality in a measurable and comparable way.

However, one must be careful to not isolate the processes and content from the context and insinuate a causal relationship between certain team behaviors and outcomes without considering the mechanics within. The limited methodology of qualitative gaming research has been conducted away from the actual gaming context, instead finding more general insights into their experiences and perceptions ~\cite{buchan2016, tan2021less}. If we wish to understand the team processes evolving throughout the game, it is important to incorporate robust real-time contextual data and give weight to individual perspectives that are unobservable from statistical or linguistic data alone.

\section{Limitations and Future Work} \label{limitations}

We acknowledge several limitations of our work. First, though our observation and interview-based study setting allowed us to better understand in-the-moment communication patterns, it could have affected participants on their in-game communication patterns. Although no participants reported any difference in how they would have behaved usually in the post-game interview beyond some discomfort with the equipment, their behaviors could have been affected. The sense of being observed by a third person can especially affect them to perform less of the behaviors that can be perceived negatively~\cite{kim2020understanding}. This could have resulted in participants not engaging in severe toxic communication patterns despite toxicity in online games being one of the frequent communication patterns~\cite{kou2020toxic, beres2021, nexo2023players}. Future work can consider using the Social Desirability Scale~\cite{mccrae1983social} to understand whether the participants are likely to behave or answer in a way that is socially desirable.

Second, we acknowledge that there could be diverse factors that could affect the result that our study did not focus on. For instance, players' demographics such as gender or nationality can affect their communication patterns. Previous work indicated that there exist gender differences in not only how players play the game but also in how they interact with other players~\cite{veltri2014gender}, such as female players are more likely to show communal attitudes or encourage others~\cite{hong2012gender}. Although our participant demographic reflects the imbalanced player demographic of \textit{LoL}, future work can investigate gender differences surrounding our three research questions. In addition, we have only recruited participants playing in the Korean \textit{LoL} server. Existing work suggests that there exist cultural differences in how \textit{LoL} players engage in toxic behaviors~\cite{sengun2019exploring}, but future work can also investigate cultural differences in other in-game communications.

We primarily focus on how players utilize and perceive the available communication features in \textit{LoL}. This work makes sense of when and why players engage or disengage from communication, weighing the risks and benefits of the communication within the confines and context of the evolving situations. Further research should look into not only when or what medium the player chooses to communicate through, but also the types of communicative behaviors, such as aggression, attribution, and socialization, to observe their impact on the game state and team communication.
\section{Conclusion}

This study sheds light on the complex communication dynamics within virtual ad hoc teams of online multiplayer games by analyzing in-the-moment communication processes of \textit{LoL} players. Our findings reveal that players navigate communication not solely through the frequency of messages but by balancing verbal and non-verbal cues in response to the demands of high-pressure gameplay and the norms established through prior experiences. Rather than viewing constant communication as inherently beneficial, players often regard action and game performance as more significant indicators of a teammate's commitment. These insights suggest that fostering effective communication in online multiplayer games requires more than just encouraging message exchange; it involves designing systems that accommodate the nuanced and context-dependent nature of player interactions. By understanding when and why players choose to communicate, as well as how their experiences shape their perceptions of their teammates, we contribute to a broader understanding of communication strategies in virtual ad hoc teams. This work points towards more tailored communication protocols that reflect the unique needs of individuals and their virtual environments, ultimately supporting more cohesive and effective team performance.

\begin{acks}
This work was supported by the Office of Naval Research (ONR: N00014-24-1-2290). We thank the \textit{League of Legends} players and community members for their insights and participation in our research.
\end{acks}

\bibliographystyle{ACM-Reference-Format}
\bibliography{main}


\begin{thebibliography}{84}


\ifx \showCODEN    \undefined \def \showCODEN     #1{\unskip}     \fi
\ifx \showDOI      \undefined \def \showDOI       #1{#1}\fi
\ifx \showISBNx    \undefined \def \showISBNx     #1{\unskip}     \fi
\ifx \showISBNxiii \undefined \def \showISBNxiii  #1{\unskip}     \fi
\ifx \showISSN     \undefined \def \showISSN      #1{\unskip}     \fi
\ifx \showLCCN     \undefined \def \showLCCN      #1{\unskip}     \fi
\ifx \shownote     \undefined \def \shownote      #1{#1}          \fi
\ifx \showarticletitle \undefined \def \showarticletitle #1{#1}   \fi
\ifx \showURL      \undefined \def \showURL       {\relax}        \fi
\providecommand\bibfield[2]{#2}
\providecommand\bibinfo[2]{#2}
\providecommand\natexlab[1]{#1}
\providecommand\showeprint[2][]{arXiv:#2}

\bibitem[Altschuller and Benbunan-Fich(2008)]%
        {altschuller2008potential}
\bibfield{author}{\bibinfo{person}{Shoshana Altschuller} {and}
  \bibinfo{person}{Raquel Benbunan-Fich}.} \bibinfo{year}{2008}\natexlab{}.
\newblock \showarticletitle{Potential antecedents to trust in ad hoc emergency
  response virtual teams}. In \bibinfo{booktitle}{\emph{Proceedings of the 5th
  International ISCRAM Conference}}. \bibinfo{pages}{254--264}.
\newblock


\bibitem[Altschuller and Benbunan-Fich(2010)]%
        {altschuller2010}
\bibfield{author}{\bibinfo{person}{Shoshana Altschuller} {and}
  \bibinfo{person}{Raquel Benbunan-Fich}.} \bibinfo{year}{2010}\natexlab{}.
\newblock \showarticletitle{Trust, performance, and the communication process
  in ad hoc decision-making virtual teams}.
\newblock \bibinfo{journal}{\emph{Journal of Computer-Mediated Communication}}
  \bibinfo{volume}{16}, \bibinfo{number}{1} (\bibinfo{year}{2010}),
  \bibinfo{pages}{27--47}.
\newblock


\bibitem[Altschuller and Benbunan-Fich(2013)]%
        {altschuller2013}
\bibfield{author}{\bibinfo{person}{Shoshana Altschuller} {and}
  \bibinfo{person}{Raquel Benbunan-Fich}.} \bibinfo{year}{2013}\natexlab{}.
\newblock \showarticletitle{The pursuit of trust in ad hoc virtual teams: how
  much electronic portrayal is too much?}
\newblock \bibinfo{journal}{\emph{European Journal of Information Systems}}
  \bibinfo{volume}{22}, \bibinfo{number}{6} (\bibinfo{year}{2013}),
  \bibinfo{pages}{619--636}.
\newblock


\bibitem[Bell et~al\mbox{.}(2016)]%
        {bell2016extreme}
\bibfield{author}{\bibinfo{person}{Suzanne Bell}, \bibinfo{person}{David
  Fisher}, \bibinfo{person}{Shanique Brown}, {and} \bibinfo{person}{Kristin
  Mann}.} \bibinfo{year}{2016}\natexlab{}.
\newblock \showarticletitle{An Approach for Conducting Actionable Research With
  Extreme Teams}.
\newblock \bibinfo{journal}{\emph{Journal of Management}}  \bibinfo{volume}{42}
  (\bibinfo{date}{06} \bibinfo{year}{2016}).
\newblock
\urldef\tempurl%
\url{https://doi.org/10.1177/0149206316653805}
\showDOI{\tempurl}


\bibitem[Beres et~al\mbox{.}(2021)]%
        {beres2021}
\bibfield{author}{\bibinfo{person}{Nicole~A Beres}, \bibinfo{person}{Julian
  Frommel}, \bibinfo{person}{Elizabeth Reid}, \bibinfo{person}{Regan~L
  Mandryk}, {and} \bibinfo{person}{Madison Klarkowski}.}
  \bibinfo{year}{2021}\natexlab{}.
\newblock \showarticletitle{Don’t You Know That You’re Toxic: Normalization
  of Toxicity in Online Gaming}. In \bibinfo{booktitle}{\emph{Proceedings of
  the 2021 CHI Conference on Human Factors in Computing Systems}} (Yokohama,
  Japan) \emph{(\bibinfo{series}{CHI '21})}. \bibinfo{publisher}{Association
  for Computing Machinery}, \bibinfo{address}{New York, NY, USA}, Article
  \bibinfo{articleno}{438}, \bibinfo{numpages}{15}~pages.
\newblock
\showISBNx{9781450380966}
\urldef\tempurl%
\url{https://doi.org/10.1145/3411764.3445157}
\showDOI{\tempurl}


\bibitem[Bisberg et~al\mbox{.}(2022)]%
        {bisberg2022contagion}
\bibfield{author}{\bibinfo{person}{Alexander~J. Bisberg},
  \bibinfo{person}{Julie Jiang}, \bibinfo{person}{Yilei Zeng},
  \bibinfo{person}{Emily Chen}, {and} \bibinfo{person}{Emilio Ferrara}.}
  \bibinfo{year}{2022}\natexlab{}.
\newblock \showarticletitle{The Gift that Keeps on Giving: Generosity is
  Contagious in Multiplayer Online Games}.
\newblock \bibinfo{journal}{\emph{Proc. ACM Hum.-Comput. Interact.}}
  \bibinfo{volume}{6}, \bibinfo{number}{CSCW2}, Article
  \bibinfo{articleno}{395} (\bibinfo{date}{Nov.} \bibinfo{year}{2022}),
  \bibinfo{numpages}{22}~pages.
\newblock
\urldef\tempurl%
\url{https://doi.org/10.1145/3555120}
\showDOI{\tempurl}


\bibitem[Braun and Clarke(2021)]%
        {braun2021saturation}
\bibfield{author}{\bibinfo{person}{Virginia Braun} {and}
  \bibinfo{person}{Victoria Clarke}.} \bibinfo{year}{2021}\natexlab{}.
\newblock \showarticletitle{To saturate or not to saturate? Questioning data
  saturation as a useful concept for thematic analysis and sample-size
  rationales}.
\newblock \bibinfo{journal}{\emph{Qualitative Research in Sport, Exercise and
  Health}} \bibinfo{volume}{13}, \bibinfo{number}{2} (\bibinfo{year}{2021}),
  \bibinfo{pages}{201--216}.
\newblock
\urldef\tempurl%
\url{https://doi.org/10.1080/2159676X.2019.1704846}
\showDOI{\tempurl}
\showeprint{https://doi.org/10.1080/2159676X.2019.1704846}


\bibitem[Br{\"u}hlmann et~al\mbox{.}(2020)]%
        {bruhlmann2020motivational}
\bibfield{author}{\bibinfo{person}{Florian Br{\"u}hlmann},
  \bibinfo{person}{Philipp Baumgartner}, \bibinfo{person}{G{\"u}nter Wallner},
  \bibinfo{person}{Simone Kriglstein}, {and} \bibinfo{person}{Elisa~D Mekler}.}
  \bibinfo{year}{2020}\natexlab{}.
\newblock \showarticletitle{Motivational profiling of league of legends
  players}.
\newblock \bibinfo{journal}{\emph{Frontiers in Psychology}}
  \bibinfo{volume}{11} (\bibinfo{year}{2020}), \bibinfo{pages}{1307}.
\newblock


\bibitem[Buchan and Taylor(2016)]%
        {buchan2016}
\bibfield{author}{\bibinfo{person}{Alexandra Buchan} {and}
  \bibinfo{person}{Jacqui Taylor}.} \bibinfo{year}{2016}\natexlab{}.
\newblock \showarticletitle{A qualitative exploration of factors affecting
  group cohesion and team play in multiplayer online battle arenas (mobas)}.
\newblock \bibinfo{journal}{\emph{The Computer Games Journal}}
  \bibinfo{volume}{5} (\bibinfo{year}{2016}), \bibinfo{pages}{65--89}.
\newblock


\bibitem[Canossa et~al\mbox{.}(2021)]%
        {canossa2021honor}
\bibfield{author}{\bibinfo{person}{Alessandro Canossa}, \bibinfo{person}{Dmitry
  Salimov}, \bibinfo{person}{Ahmad Azadvar}, \bibinfo{person}{Casper
  Harteveld}, {and} \bibinfo{person}{Georgios Yannakakis}.}
  \bibinfo{year}{2021}\natexlab{}.
\newblock \showarticletitle{For Honor, for Toxicity: Detecting Toxic Behavior
  through Gameplay}.
\newblock \bibinfo{journal}{\emph{Proc. ACM Hum.-Comput. Interact.}}
  \bibinfo{volume}{5}, \bibinfo{number}{CHI PLAY}, Article
  \bibinfo{articleno}{253} (\bibinfo{date}{Oct.} \bibinfo{year}{2021}),
  \bibinfo{numpages}{29}~pages.
\newblock
\urldef\tempurl%
\url{https://doi.org/10.1145/3474680}
\showDOI{\tempurl}


\bibitem[Capiola et~al\mbox{.}(2020)]%
        {capiola2020}
\bibfield{author}{\bibinfo{person}{August Capiola}, \bibinfo{person}{Holly~C.
  Baxter}, \bibinfo{person}{Marc~D. Pfahler}, \bibinfo{person}{Christopher~S.
  Calhoun}, {and} \bibinfo{person}{Philip Bobko}.}
  \bibinfo{year}{2020}\natexlab{}.
\newblock \showarticletitle{Swift Trust in Ad Hoc Teams: A Cognitive Task
  Analysis of Intelligence Operators in Multi-Domain Command and Control
  Contexts}.
\newblock \bibinfo{journal}{\emph{Journal of Cognitive Engineering and Decision
  Making}} \bibinfo{volume}{14}, \bibinfo{number}{3} (\bibinfo{year}{2020}),
  \bibinfo{pages}{218--241}.
\newblock
\urldef\tempurl%
\url{https://doi.org/10.1177/1555343420943460}
\showDOI{\tempurl}


\bibitem[Chalupnik and Atkins(2020)]%
        {chalupnik2020}
\bibfield{author}{\bibinfo{person}{Malgorzata Chalupnik} {and}
  \bibinfo{person}{Sarah Atkins}.} \bibinfo{year}{2020}\natexlab{}.
\newblock \bibinfo{title}{Team interaction in healthcare settings: Leadership,
  rapport-building and clinical outcomes in ad hoc medical teams.}
\newblock , \bibinfo{numpages}{55-83}~pages.
\newblock
\urldef\tempurl%
\url{https://doi.org/10.1075/pbns.311}
\showDOI{\tempurl}


\bibitem[Cherry and Robillard(2008)]%
        {cherry2008}
\bibfield{author}{\bibinfo{person}{S{\'e}bastien Cherry} {and}
  \bibinfo{person}{Pierre~N Robillard}.} \bibinfo{year}{2008}\natexlab{}.
\newblock \showarticletitle{The social side of software engineering—A real ad
  hoc collaboration network}.
\newblock \bibinfo{journal}{\emph{International Journal of Human-Computer
  Studies}} \bibinfo{volume}{66}, \bibinfo{number}{7} (\bibinfo{year}{2008}),
  \bibinfo{pages}{495--505}.
\newblock


\bibitem[Cote(2017)]%
        {cote2017coping}
\bibfield{author}{\bibinfo{person}{Amanda Cote}.}
  \bibinfo{year}{2017}\natexlab{}.
\newblock \showarticletitle{"I Can Defend Myself": Women's Strategies for
  Coping With Harassment While Gaming Online}.
\newblock \bibinfo{journal}{\emph{Games and Culture}}  \bibinfo{volume}{12}
  (\bibinfo{date}{03} \bibinfo{year}{2017}), \bibinfo{pages}{136--155}.
\newblock
\urldef\tempurl%
\url{https://doi.org/10.1177/1555412015587603}
\showDOI{\tempurl}


\bibitem[Deslauriers et~al\mbox{.}(2020)]%
        {deslauriers2020dbd}
\bibfield{author}{\bibinfo{person}{Patrick Deslauriers},
  \bibinfo{person}{Laura~Iseut Lafrance St-Martin}, {and}
  \bibinfo{person}{Maude Bonenfant}.} \bibinfo{year}{2020}\natexlab{}.
\newblock \showarticletitle{Assessing Toxic Behaviour in Dead by Daylight
  Perceptions and Factors of Toxicity According to the Game’s Official
  Subreddit Contributors}.
\newblock \bibinfo{journal}{\emph{Game Studies}}  \bibinfo{volume}{20}
  (\bibinfo{date}{12} \bibinfo{year}{2020}).
\newblock


\bibitem[Dirks and de~Jong(2022)]%
        {dirks2022trust}
\bibfield{author}{\bibinfo{person}{Kurt~T. Dirks} {and} \bibinfo{person}{Bart
  de Jong}.} \bibinfo{year}{2022}\natexlab{}.
\newblock \showarticletitle{Trust Within the Workplace: A Review of Two Waves
  of Research and a Glimpse of the Third}.
\newblock \bibinfo{journal}{\emph{Annual Review of Organizational Psychology
  and Organizational Behavior}} \bibinfo{volume}{9}, \bibinfo{number}{Volume 9,
  2022} (\bibinfo{year}{2022}), \bibinfo{pages}{247--276}.
\newblock
\showISSN{2327-0616}
\urldef\tempurl%
\url{https://doi.org/10.1146/annurev-orgpsych-012420-083025}
\showDOI{\tempurl}


\bibitem[Eisenberg and Krishnan(2018)]%
        {eisenberg2018}
\bibfield{author}{\bibinfo{person}{Julia Eisenberg} {and}
  \bibinfo{person}{Aparna Krishnan}.} \bibinfo{year}{2018}\natexlab{}.
\newblock \showarticletitle{Addressing Virtual Work Challenges: Learning From
  the Field}.
\newblock \bibinfo{journal}{\emph{Organization Management Journal}}
  \bibinfo{volume}{15}, \bibinfo{number}{2} (\bibinfo{year}{2018}),
  \bibinfo{pages}{78--94}.
\newblock
\urldef\tempurl%
\url{https://doi.org/10.1080/15416518.2018.1471976}
\showDOI{\tempurl}
\showeprint{https://doi.org/10.1080/15416518.2018.1471976}


\bibitem[Elo and Kyng{\"a}s(2008)]%
        {elo2008qualitative}
\bibfield{author}{\bibinfo{person}{Satu Elo} {and} \bibinfo{person}{Helvi
  Kyng{\"a}s}.} \bibinfo{year}{2008}\natexlab{}.
\newblock \showarticletitle{The qualitative content analysis process}.
\newblock \bibinfo{journal}{\emph{Journal of advanced nursing}}
  \bibinfo{volume}{62}, \bibinfo{number}{1} (\bibinfo{year}{2008}),
  \bibinfo{pages}{107--115}.
\newblock


\bibitem[Entin and Serfaty(1999)]%
        {entin1999}
\bibfield{author}{\bibinfo{person}{Elliot~E Entin} {and}
  \bibinfo{person}{Daniel Serfaty}.} \bibinfo{year}{1999}\natexlab{}.
\newblock \showarticletitle{Adaptive team coordination}.
\newblock \bibinfo{journal}{\emph{Human factors}} \bibinfo{volume}{41},
  \bibinfo{number}{2} (\bibinfo{year}{1999}), \bibinfo{pages}{312--325}.
\newblock


\bibitem[Erdem and Özen Aytemur(2003)]%
        {erdem2003cognitive}
\bibfield{author}{\bibinfo{person}{Ferda Erdem} {and} \bibinfo{person}{Janset
  Özen Aytemur}.} \bibinfo{year}{2003}\natexlab{}.
\newblock \showarticletitle{Cognitive and Affective Dimensions of Trust in
  Developing Team Performance}.
\newblock \bibinfo{journal}{\emph{Team Performance Management}}
  \bibinfo{volume}{9} (\bibinfo{date}{09} \bibinfo{year}{2003}),
  \bibinfo{pages}{131--135}.
\newblock
\urldef\tempurl%
\url{https://doi.org/10.1108/13527590310493846}
\showDOI{\tempurl}


\bibitem[Evans et~al\mbox{.}(2021)]%
        {evans2021}
\bibfield{author}{\bibinfo{person}{J~Colin Evans}, \bibinfo{person}{M~Blair
  Evans}, \bibinfo{person}{Meagan Slack}, \bibinfo{person}{Michael Peddle},
  {and} \bibinfo{person}{Lorelei Lingard}.} \bibinfo{year}{2021}\natexlab{}.
\newblock \showarticletitle{Examining non-technical skills for ad hoc
  resuscitation teams: a scoping review and taxonomy of team-related concepts}.
\newblock \bibinfo{journal}{\emph{Scandinavian journal of trauma, resuscitation
  and emergency medicine}}  \bibinfo{volume}{29} (\bibinfo{year}{2021}),
  \bibinfo{pages}{1--22}.
\newblock


\bibitem[Finholt et~al\mbox{.}(1990)]%
        {finholt1990}
\bibfield{author}{\bibinfo{person}{Tom Finholt}, \bibinfo{person}{Lee Sproull},
  {and} \bibinfo{person}{Sara Kiesler}.} \bibinfo{year}{1990}\natexlab{}.
\newblock \showarticletitle{Communication and performance in ad hoc task
  groups}.
\newblock \bibinfo{journal}{\emph{Intellectual teamwork}}
  (\bibinfo{year}{1990}), \bibinfo{pages}{291--325}.
\newblock


\bibitem[Fisher(2023)]%
        {carver2023}
\bibfield{author}{\bibinfo{person}{Carver Fisher}.}
  \bibinfo{year}{2023}\natexlab{}.
\newblock
\newblock
\urldef\tempurl%
\url{https://www.dexerto.com/league-of-legends/riot-dev-explains-why-league-of-legends-doesnt-have-voice-chat-like-valorant-2079910/}
\showURL{%
\tempurl}


\bibitem[Fox and Tang(2016)]%
        {fox2016women}
\bibfield{author}{\bibinfo{person}{Jesse Fox} {and} \bibinfo{person}{Wai~Yen
  Tang}.} \bibinfo{year}{2016}\natexlab{}.
\newblock \showarticletitle{Women's experiences with general and sexual
  harassment in online video games: Rumination, organizational responsiveness,
  withdrawal, and coping strategies}.
\newblock \bibinfo{journal}{\emph{New Media \& Society}}  \bibinfo{volume}{19}
  (\bibinfo{date}{03} \bibinfo{year}{2016}).
\newblock
\urldef\tempurl%
\url{https://doi.org/10.1177/1461444816635778}
\showDOI{\tempurl}


\bibitem[Grimm(2010)]%
        {grimm2010social}
\bibfield{author}{\bibinfo{person}{Pamela Grimm}.}
  \bibinfo{year}{2010}\natexlab{}.
\newblock \bibinfo{booktitle}{\emph{Social Desirability Bias}}.
\newblock \bibinfo{publisher}{John Wiley \& Sons, Ltd}.
\newblock
\showISBNx{9781444316568}
\urldef\tempurl%
\url{https://doi.org/10.1002/9781444316568.wiem02057}
\showDOI{\tempurl}
\showeprint{https://onlinelibrary.wiley.com/doi/pdf/10.1002/9781444316568.wiem02057}


\bibitem[Harrison and Connors(1984)]%
        {harrison1984exotic}
\bibfield{author}{\bibinfo{person}{Albert~A. Harrison} {and}
  \bibinfo{person}{Mary~M. Connors}.} \bibinfo{year}{1984}\natexlab{}.
\newblock \showarticletitle{Groups in Exotic Environments}.
\newblock \bibinfo{series}{Advances in Experimental Social Psychology},
  Vol.~\bibinfo{volume}{18}. \bibinfo{publisher}{Academic Press},
  \bibinfo{pages}{49--87}.
\newblock
\showISSN{0065-2601}
\urldef\tempurl%
\url{https://doi.org/10.1016/S0065-2601(08)60142-2}
\showDOI{\tempurl}


\bibitem[Hong and Hwang(2012)]%
        {hong2012gender}
\bibfield{author}{\bibinfo{person}{Jon-Chao Hong} {and}
  \bibinfo{person}{Ming-Yueh Hwang}.} \bibinfo{year}{2012}\natexlab{}.
\newblock \showarticletitle{Gender differences in help-seeking and supportive
  dialogue during on-line game}.
\newblock \bibinfo{journal}{\emph{Procedia-Social and Behavioral Sciences}}
  \bibinfo{volume}{64} (\bibinfo{year}{2012}), \bibinfo{pages}{342--351}.
\newblock


\bibitem[Jarvenpaa and Leidner(1998)]%
        {jarvenpaa1998}
\bibfield{author}{\bibinfo{person}{Sirkka~L. Jarvenpaa} {and}
  \bibinfo{person}{Dorothy~E. Leidner}.} \bibinfo{year}{1998}\natexlab{}.
\newblock \showarticletitle{{Communication and Trust in Global Virtual Teams}}.
\newblock \bibinfo{journal}{\emph{Journal of Computer-Mediated Communication}}
  \bibinfo{volume}{3}, \bibinfo{number}{4} (\bibinfo{date}{06}
  \bibinfo{year}{1998}), \bibinfo{pages}{JCMC346}.
\newblock
\urldef\tempurl%
\url{https://doi.org/10.1111/j.1083-6101.1998.tb00080.x}
\showDOI{\tempurl}


\bibitem[Kanawattanachai and Yoo(2002)]%
        {kanawattanachai2002dynamic}
\bibfield{author}{\bibinfo{person}{Prasert Kanawattanachai} {and}
  \bibinfo{person}{Youngjin Yoo}.} \bibinfo{year}{2002}\natexlab{}.
\newblock \showarticletitle{Dynamic Nature of Trust in Virtual Teams}.
\newblock \bibinfo{journal}{\emph{The Journal of Strategic Information
  Systems}}  \bibinfo{volume}{11} (\bibinfo{date}{12} \bibinfo{year}{2002}),
  \bibinfo{pages}{187--213}.
\newblock
\urldef\tempurl%
\url{https://doi.org/10.1016/S0963-8687(02)00019-7}
\showDOI{\tempurl}


\bibitem[Kim et~al\mbox{.}(2020)]%
        {kim2020understanding}
\bibfield{author}{\bibinfo{person}{Seoyoung Kim}, \bibinfo{person}{Arti
  Thakur}, {and} \bibinfo{person}{Juho Kim}.} \bibinfo{year}{2020}\natexlab{}.
\newblock \showarticletitle{Understanding Users' Perception Towards Automated
  Personality Detection with Group-specific Behavioral Data}. In
  \bibinfo{booktitle}{\emph{Proceedings of the 2020 CHI Conference on Human
  Factors in Computing Systems}}. \bibinfo{pages}{1--12}.
\newblock


\bibitem[Kim et~al\mbox{.}(2017)]%
        {kim2017}
\bibfield{author}{\bibinfo{person}{Young~Ji Kim}, \bibinfo{person}{David
  Engel}, \bibinfo{person}{Anita~Williams Woolley}, \bibinfo{person}{Jeffrey
  Yu-Ting Lin}, \bibinfo{person}{Naomi McArthur}, {and}
  \bibinfo{person}{Thomas~W. Malone}.} \bibinfo{year}{2017}\natexlab{}.
\newblock \showarticletitle{What Makes a Strong Team? Using Collective
  Intelligence to Predict Team Performance in League of Legends}. In
  \bibinfo{booktitle}{\emph{Proceedings of the 2017 ACM Conference on Computer
  Supported Cooperative Work and Social Computing}} (Portland, Oregon, USA)
  \emph{(\bibinfo{series}{CSCW '17})}. \bibinfo{publisher}{Association for
  Computing Machinery}, \bibinfo{address}{New York, NY, USA},
  \bibinfo{pages}{2316–2329}.
\newblock
\showISBNx{9781450343350}
\urldef\tempurl%
\url{https://doi.org/10.1145/2998181.2998185}
\showDOI{\tempurl}


\bibitem[Kordyaka et~al\mbox{.}(2023)]%
        {kordyaka2023gender}
\bibfield{author}{\bibinfo{person}{Bastian Kordyaka}, \bibinfo{person}{Luisa
  Pumplun}, \bibinfo{person}{Marlies Brunnhofer}, \bibinfo{person}{Bjoern
  Kruse}, {and} \bibinfo{person}{Samuli Laato}.}
  \bibinfo{year}{2023}\natexlab{}.
\newblock \showarticletitle{Gender disparities in esports--An explanatory
  mixed-methods approach}.
\newblock \bibinfo{journal}{\emph{Computers in Human Behavior}}
  \bibinfo{volume}{149} (\bibinfo{year}{2023}), \bibinfo{pages}{107956}.
\newblock


\bibitem[Kou(2020)]%
        {kou2020toxic}
\bibfield{author}{\bibinfo{person}{Yubo Kou}.} \bibinfo{year}{2020}\natexlab{}.
\newblock \showarticletitle{Toxic Behaviors in Team-Based Competitive Gaming:
  The Case of League of Legends}. In \bibinfo{booktitle}{\emph{Proceedings of
  the Annual Symposium on Computer-Human Interaction in Play}} (Virtual Event,
  Canada) \emph{(\bibinfo{series}{CHI PLAY '20})}.
  \bibinfo{publisher}{Association for Computing Machinery},
  \bibinfo{address}{New York, NY, USA}, \bibinfo{pages}{81–92}.
\newblock
\showISBNx{9781450380744}
\urldef\tempurl%
\url{https://doi.org/10.1145/3410404.3414243}
\showDOI{\tempurl}


\bibitem[Kou and Gui(2014)]%
        {kou2014}
\bibfield{author}{\bibinfo{person}{Yubo Kou} {and} \bibinfo{person}{Xinning
  Gui}.} \bibinfo{year}{2014}\natexlab{}.
\newblock \showarticletitle{Playing with strangers: understanding temporary
  teams in league of legends}. In \bibinfo{booktitle}{\emph{Proceedings of the
  First ACM SIGCHI Annual Symposium on Computer-Human Interaction in Play}}
  (Toronto, Ontario, Canada) \emph{(\bibinfo{series}{CHI PLAY '14})}.
  \bibinfo{publisher}{Association for Computing Machinery},
  \bibinfo{address}{New York, NY, USA}, \bibinfo{pages}{161–169}.
\newblock
\showISBNx{9781450330145}
\urldef\tempurl%
\url{https://doi.org/10.1145/2658537.2658538}
\showDOI{\tempurl}


\bibitem[Kou and Gui(2018)]%
        {kou2018self}
\bibfield{author}{\bibinfo{person}{Yubo Kou} {and} \bibinfo{person}{Xinning
  Gui}.} \bibinfo{year}{2018}\natexlab{}.
\newblock \showarticletitle{Entangled with Numbers: Quantified Self and Others
  in a Team-Based Online Game}.
\newblock \bibinfo{journal}{\emph{Proc. ACM Hum.-Comput. Interact.}}
  \bibinfo{volume}{2}, \bibinfo{number}{CSCW}, Article \bibinfo{articleno}{93}
  (\bibinfo{date}{Nov.} \bibinfo{year}{2018}), \bibinfo{numpages}{25}~pages.
\newblock
\urldef\tempurl%
\url{https://doi.org/10.1145/3274362}
\showDOI{\tempurl}


\bibitem[Kwak and Blackburn(2015)]%
        {kwak2015linguistic}
\bibfield{author}{\bibinfo{person}{Haewoon Kwak} {and} \bibinfo{person}{Jeremy
  Blackburn}.} \bibinfo{year}{2015}\natexlab{}.
\newblock \showarticletitle{Linguistic Analysis of Toxic Behavior in an Online
  Video Game}. In \bibinfo{booktitle}{\emph{Social Informatics: SocInfo 2014
  International Workshops, Barcelona, Spain, November 11, 2014, Revised
  Selected Papers}} (Barcelona, Spain). \bibinfo{publisher}{Springer-Verlag},
  \bibinfo{address}{Berlin, Heidelberg}, \bibinfo{pages}{209–217}.
\newblock
\showISBNx{978-3-319-15167-0}
\urldef\tempurl%
\url{https://doi.org/10.1007/978-3-319-15168-7_26}
\showDOI{\tempurl}


\bibitem[Kwak et~al\mbox{.}(2015)]%
        {kwak2015exploring}
\bibfield{author}{\bibinfo{person}{Haewoon Kwak}, \bibinfo{person}{Jeremy
  Blackburn}, {and} \bibinfo{person}{Seungyeop Han}.}
  \bibinfo{year}{2015}\natexlab{}.
\newblock \showarticletitle{Exploring Cyberbullying and Other Toxic Behavior in
  Team Competition Online Games}. In \bibinfo{booktitle}{\emph{Proceedings of
  the 33rd Annual ACM Conference on Human Factors in Computing Systems}}
  (Seoul, Republic of Korea) \emph{(\bibinfo{series}{CHI '15})}.
  \bibinfo{publisher}{Association for Computing Machinery},
  \bibinfo{address}{New York, NY, USA}, \bibinfo{pages}{3739–3748}.
\newblock
\showISBNx{9781450331456}
\urldef\tempurl%
\url{https://doi.org/10.1145/2702123.2702529}
\showDOI{\tempurl}


\bibitem[Laato et~al\mbox{.}(2024)]%
        {laato2024starcraft}
\bibfield{author}{\bibinfo{person}{Samuli Laato}, \bibinfo{person}{Bastian
  Kordyaka}, {and} \bibinfo{person}{Juho Hamari}.}
  \bibinfo{year}{2024}\natexlab{}.
\newblock \showarticletitle{Traumatizing or Just Annoying? Unveiling the
  Spectrum of Gamer Toxicity in the StarCraft II Community}. In
  \bibinfo{booktitle}{\emph{Proceedings of the 2024 CHI Conference on Human
  Factors in Computing Systems}} (Honolulu, HI, USA)
  \emph{(\bibinfo{series}{CHI '24})}. \bibinfo{publisher}{Association for
  Computing Machinery}, \bibinfo{address}{New York, NY, USA}, Article
  \bibinfo{articleno}{758}, \bibinfo{numpages}{18}~pages.
\newblock
\showISBNx{9798400703300}
\urldef\tempurl%
\url{https://doi.org/10.1145/3613904.3642137}
\showDOI{\tempurl}


\bibitem[Leavitt et~al\mbox{.}(2016)]%
        {leavitt2016}
\bibfield{author}{\bibinfo{person}{Alex Leavitt}, \bibinfo{person}{Brian~C.
  Keegan}, {and} \bibinfo{person}{Joshua Clark}.}
  \bibinfo{year}{2016}\natexlab{}.
\newblock \showarticletitle{Ping to Win? Non-Verbal Communication and Team
  Performance in Competitive Online Multiplayer Games}. In
  \bibinfo{booktitle}{\emph{Proceedings of the 2016 CHI Conference on Human
  Factors in Computing Systems}} (San Jose, California, USA)
  \emph{(\bibinfo{series}{CHI '16})}. \bibinfo{publisher}{Association for
  Computing Machinery}, \bibinfo{address}{New York, NY, USA},
  \bibinfo{pages}{4337–4350}.
\newblock
\showISBNx{9781450333627}
\urldef\tempurl%
\url{https://doi.org/10.1145/2858036.2858132}
\showDOI{\tempurl}


\bibitem[Lines et~al\mbox{.}(2022)]%
        {lines2022meta}
\bibfield{author}{\bibinfo{person}{Robin Lines}, \bibinfo{person}{Michael
  Chapman}, \bibinfo{person}{Leo Ng}, \bibinfo{person}{Sasha Nahleen},
  \bibinfo{person}{Philip Temby}, \bibinfo{person}{Monique Crane},
  \bibinfo{person}{Gavin Hazel}, {and} \bibinfo{person}{F. Gucciardi}.}
  \bibinfo{year}{2022}\natexlab{}.
\newblock \showarticletitle{A Meta-Analytic Test of Trust Formation and
  Development in Swift Starting Action Teams}.
\newblock \bibinfo{journal}{\emph{Sport, Exercise, and Performance Psychology}}
   \bibinfo{volume}{11} (\bibinfo{date}{06} \bibinfo{year}{2022}).
\newblock
\urldef\tempurl%
\url{https://doi.org/10.1037/spy0000298}
\showDOI{\tempurl}


\bibitem[Longstaff and Yang(2008)]%
        {longstaff2008natural}
\bibfield{author}{\bibinfo{person}{P. Longstaff} {and} \bibinfo{person}{Sung-Un
  Yang}.} \bibinfo{year}{2008}\natexlab{}.
\newblock \showarticletitle{Communication Management and Trust: Their Role in
  Building Resilience to 'Surprises' Such As Natural Disasters, Pandemic Flu,
  and Terrorism}.
\newblock \bibinfo{journal}{\emph{Ecology and Society}}  \bibinfo{volume}{1}
  (\bibinfo{date}{06} \bibinfo{year}{2008}).
\newblock
\urldef\tempurl%
\url{https://doi.org/10.5751/ES-02232-130103}
\showDOI{\tempurl}


\bibitem[Marlow et~al\mbox{.}(2018)]%
        {marlow2018}
\bibfield{author}{\bibinfo{person}{Shannon~L Marlow},
  \bibinfo{person}{Christina~N Lacerenza}, \bibinfo{person}{Jensine Paoletti},
  \bibinfo{person}{C~Shawn Burke}, {and} \bibinfo{person}{Eduardo Salas}.}
  \bibinfo{year}{2018}\natexlab{}.
\newblock \showarticletitle{Does team communication represent a
  one-size-fits-all approach?: A meta-analysis of team communication and
  performance}.
\newblock \bibinfo{journal}{\emph{Organizational behavior and human decision
  processes}}  \bibinfo{volume}{144} (\bibinfo{year}{2018}),
  \bibinfo{pages}{145--170}.
\newblock


\bibitem[Marlow et~al\mbox{.}(2017)]%
        {marlow2017}
\bibfield{author}{\bibinfo{person}{Shannon~L Marlow},
  \bibinfo{person}{Christina~N Lacerenza}, {and} \bibinfo{person}{Eduardo
  Salas}.} \bibinfo{year}{2017}\natexlab{}.
\newblock \showarticletitle{Communication in virtual teams: A conceptual
  framework and research agenda}.
\newblock \bibinfo{journal}{\emph{Human resource management review}}
  \bibinfo{volume}{27}, \bibinfo{number}{4} (\bibinfo{year}{2017}),
  \bibinfo{pages}{575--589}.
\newblock


\bibitem[Mayer et~al\mbox{.}(1995)]%
        {mayer1995trust}
\bibfield{author}{\bibinfo{person}{Roger~C. Mayer}, \bibinfo{person}{James~H.
  Davis}, {and} \bibinfo{person}{F.~David Schoorman}.}
  \bibinfo{year}{1995}\natexlab{}.
\newblock \showarticletitle{An Integrative Model of Organizational Trust}.
\newblock \bibinfo{journal}{\emph{The Academy of Management Review}}
  \bibinfo{volume}{20}, \bibinfo{number}{3} (\bibinfo{year}{1995}),
  \bibinfo{pages}{709--734}.
\newblock
\showISSN{03637425}
\urldef\tempurl%
\url{http://www.jstor.org/stable/258792}
\showURL{%
\tempurl}


\bibitem[McCrae and Costa(1983)]%
        {mccrae1983social}
\bibfield{author}{\bibinfo{person}{Robert~R McCrae} {and}
  \bibinfo{person}{Paul~T Costa}.} \bibinfo{year}{1983}\natexlab{}.
\newblock \showarticletitle{Social desirability scales: More substance than
  style.}
\newblock \bibinfo{journal}{\emph{Journal of consulting and clinical
  psychology}} \bibinfo{volume}{51}, \bibinfo{number}{6}
  (\bibinfo{year}{1983}), \bibinfo{pages}{882}.
\newblock


\bibitem[McLean and Griffiths(2019)]%
        {mclean2019female}
\bibfield{author}{\bibinfo{person}{Lavinia McLean} {and}
  \bibinfo{person}{Mark~D Griffiths}.} \bibinfo{year}{2019}\natexlab{}.
\newblock \showarticletitle{Female gamers’ experience of online harassment
  and social support in online gaming: A qualitative study}.
\newblock \bibinfo{journal}{\emph{International Journal of Mental Health and
  Addiction}}  \bibinfo{volume}{17} (\bibinfo{year}{2019}),
  \bibinfo{pages}{970--994}.
\newblock


\bibitem[Mesmer-Magnus and DeChurch(2009)]%
        {mesmer2009}
\bibfield{author}{\bibinfo{person}{Jessica~R Mesmer-Magnus} {and}
  \bibinfo{person}{Leslie~A DeChurch}.} \bibinfo{year}{2009}\natexlab{}.
\newblock \showarticletitle{Information sharing and team performance: a
  meta-analysis.}
\newblock \bibinfo{journal}{\emph{Journal of applied psychology}}
  \bibinfo{volume}{94}, \bibinfo{number}{2} (\bibinfo{year}{2009}),
  \bibinfo{pages}{535}.
\newblock


\bibitem[Monge and O’Brien(2022)]%
        {monge2022effects}
\bibfield{author}{\bibinfo{person}{CK Monge} {and} \bibinfo{person}{TC
  O’Brien}.} \bibinfo{year}{2022}\natexlab{}.
\newblock \showarticletitle{Effects of individual toxic behavior on team
  performance in League of Legends}.
\newblock \bibinfo{journal}{\emph{Media Psychology}} \bibinfo{volume}{25},
  \bibinfo{number}{1} (\bibinfo{year}{2022}), \bibinfo{pages}{82--105}.
\newblock


\bibitem[Monge and Contractor(2003)]%
        {monge2003}
\bibfield{author}{\bibinfo{person}{Peter~R Monge} {and}
  \bibinfo{person}{Noshir~S Contractor}.} \bibinfo{year}{2003}\natexlab{}.
\newblock \bibinfo{booktitle}{\emph{Theories of communication networks}}.
\newblock \bibinfo{publisher}{Oxford University Press, USA}.
\newblock


\bibitem[Morrison-Smith and Ruiz(2020)]%
        {morrison2020}
\bibfield{author}{\bibinfo{person}{Sarah Morrison-Smith} {and}
  \bibinfo{person}{Jaime Ruiz}.} \bibinfo{year}{2020}\natexlab{}.
\newblock \showarticletitle{Challenges and barriers in virtual teams: a
  literature review}.
\newblock \bibinfo{journal}{\emph{SN Applied Sciences}} \bibinfo{volume}{2},
  \bibinfo{number}{6} (\bibinfo{year}{2020}), \bibinfo{pages}{1--33}.
\newblock


\bibitem[Mun(2019)]%
        {kukinews2019}
\bibfield{author}{\bibinfo{person}{Daechan Mun}.} \bibinfo{year}{October 17,
  2019}\natexlab{}.
\newblock \bibinfo{title}{[Feature] How 'LoL' Became a 'National Game' in
  Korea}.
\newblock
  \bibinfo{howpublished}{\url{https://m.kukinews.com/article/view/kuk201910160358\#_enliple}}.
\newblock


\bibitem[Musick et~al\mbox{.}(2021)]%
        {musick2021cognition}
\bibfield{author}{\bibinfo{person}{Geoff Musick}, \bibinfo{person}{Rui Zhang},
  \bibinfo{person}{Nathan McNeese}, \bibinfo{person}{Guo Freeman}, {and}
  \bibinfo{person}{Anurata Hridi}.} \bibinfo{year}{2021}\natexlab{}.
\newblock \showarticletitle{Leveling Up Teamwork in Esports: Understanding Team
  Cognition in a Dynamic Virtual Environment}.
\newblock \bibinfo{journal}{\emph{Proceedings of the ACM on Human-Computer
  Interaction}}  \bibinfo{volume}{5} (\bibinfo{date}{04} \bibinfo{year}{2021}),
  \bibinfo{pages}{1--30}.
\newblock
\urldef\tempurl%
\url{https://doi.org/10.1145/3449123}
\showDOI{\tempurl}


\bibitem[Nex{\o} and Kristiansen(2023)]%
        {nexo2023players}
\bibfield{author}{\bibinfo{person}{Louise~Anker Nex{\o}} {and}
  \bibinfo{person}{S{\o}ren Kristiansen}.} \bibinfo{year}{2023}\natexlab{}.
\newblock \showarticletitle{Players don’t die, they Respawn: A situational
  analysis of toxic encounters arising from death events in League of Legends}.
\newblock \bibinfo{journal}{\emph{European Journal on Criminal Policy and
  Research}} \bibinfo{volume}{29}, \bibinfo{number}{3} (\bibinfo{year}{2023}),
  \bibinfo{pages}{457--476}.
\newblock


\bibitem[Norris(2004)]%
        {norris2004gender}
\bibfield{author}{\bibinfo{person}{Kamala~O Norris}.}
  \bibinfo{year}{2004}\natexlab{}.
\newblock \showarticletitle{Gender stereotypes, aggression, and computer games:
  An online survey of women}.
\newblock \bibinfo{journal}{\emph{Cyberpsychology \& Behavior}}
  \bibinfo{volume}{7}, \bibinfo{number}{6} (\bibinfo{year}{2004}),
  \bibinfo{pages}{714--727}.
\newblock


\bibitem[of~Graphs(2024a)]%
        {log2024ping}
\bibfield{author}{\bibinfo{person}{League of Graphs}.}
  \bibinfo{year}{2024}\natexlab{a}.
\newblock \bibinfo{title}{Pings stats}.
\newblock
\newblock
\urldef\tempurl%
\url{https://www.leagueofgraphs.com/stats/pings/kr/iron}
\showURL{%
\tempurl}


\bibitem[of~Graphs(2024b)]%
        {log2024}
\bibfield{author}{\bibinfo{person}{League of Graphs}.}
  \bibinfo{year}{2024}\natexlab{b}.
\newblock \bibinfo{title}{Rank distribution, KR server}.
\newblock
\newblock
\urldef\tempurl%
\url{https://www.leagueofgraphs.com/rankings/rank-distribution/kr}
\showURL{%
\tempurl}


\bibitem[Oh(2024)]%
        {newsis2024}
\bibfield{author}{\bibinfo{person}{Donghyeon Oh}.} \bibinfo{year}{October 10,
  2024}\natexlab{}.
\newblock \bibinfo{title}{7 out of 10 teenage boys in Korea play 'LoL'... 'Lee
  Sin' is popular. (translated)}.
\newblock
  \bibinfo{howpublished}{\url{https://www.newsis.com/view/NISX20241010_0002914853}}.
\newblock


\bibitem[O'Leary and Cummings(2007)]%
        {oleary2007}
\bibfield{author}{\bibinfo{person}{Michael~Boyer O'Leary} {and}
  \bibinfo{person}{Jonathon~N Cummings}.} \bibinfo{year}{2007}\natexlab{}.
\newblock \showarticletitle{The spatial, temporal, and configurational
  characteristics of geographic dispersion in teams}.
\newblock \bibinfo{journal}{\emph{MIS quarterly}} (\bibinfo{year}{2007}),
  \bibinfo{pages}{433--452}.
\newblock


\bibitem[Ong et~al\mbox{.}(2015)]%
        {ong2015player}
\bibfield{author}{\bibinfo{person}{Hao~Yi Ong}, \bibinfo{person}{Sunil
  Deolalikar}, {and} \bibinfo{person}{Mark Peng}.}
  \bibinfo{year}{2015}\natexlab{}.
\newblock \showarticletitle{Player behavior and optimal team composition for
  online multiplayer games}.
\newblock \bibinfo{journal}{\emph{arXiv preprint arXiv:1503.02230}}
  (\bibinfo{year}{2015}).
\newblock


\bibitem[Pascual et~al\mbox{.}(1999)]%
        {pascual1999}
\bibfield{author}{\bibinfo{person}{RG Pascual}, \bibinfo{person}{MC Mills},
  {and} \bibinfo{person}{Carol Blendell}.} \bibinfo{year}{1999}\natexlab{}.
\newblock \showarticletitle{Supporting distributed and ad-hoc team
  interaction}. In \bibinfo{booktitle}{\emph{1999 International Conference on
  Human Interfaces in Control Rooms, Cockpits and Command Centres}}. IET,
  \bibinfo{pages}{64--71}.
\newblock


\bibitem[Pe{\~n}arroja et~al\mbox{.}(2013)]%
        {penarroja2013}
\bibfield{author}{\bibinfo{person}{Vicente Pe{\~n}arroja},
  \bibinfo{person}{Virginia Orengo}, \bibinfo{person}{Ana Zornoza}, {and}
  \bibinfo{person}{Ana Hern{\'a}ndez}.} \bibinfo{year}{2013}\natexlab{}.
\newblock \showarticletitle{The effects of virtuality level on task-related
  collaborative behaviors: The mediating role of team trust}.
\newblock \bibinfo{journal}{\emph{Computers in Human Behavior}}
  \bibinfo{volume}{29}, \bibinfo{number}{3} (\bibinfo{year}{2013}),
  \bibinfo{pages}{967--974}.
\newblock


\bibitem[Petter and Petter(2011)]%
        {petter2011}
\bibfield{author}{\bibinfo{person}{Stacie Petter} {and} \bibinfo{person}{Rick
  Petter}.} \bibinfo{year}{2011}\natexlab{}.
\newblock \showarticletitle{To Tell or Not to Tell: Examining Team Silence and
  Voice in Online Ad Hoc Teams.}
\newblock \bibinfo{journal}{\emph{International Conference on Information
  Systems 2011, ICIS 2011}}  \bibinfo{volume}{4} (\bibinfo{date}{01}
  \bibinfo{year}{2011}).
\newblock


\bibitem[Poeller et~al\mbox{.}(2023)]%
        {poeller2023}
\bibfield{author}{\bibinfo{person}{Susanne Poeller},
  \bibinfo{person}{Martin~Johannes Dechant}, \bibinfo{person}{Madison
  Klarkowski}, {and} \bibinfo{person}{Regan~L. Mandryk}.}
  \bibinfo{year}{2023}\natexlab{}.
\newblock \showarticletitle{Suspecting Sarcasm: How League of Legends Players
  Dismiss Positive Communication in Toxic Environments}.
\newblock \bibinfo{journal}{\emph{Proc. ACM Hum.-Comput. Interact.}}
  \bibinfo{volume}{7}, \bibinfo{number}{CHI PLAY}, Article
  \bibinfo{articleno}{374} (\bibinfo{date}{oct} \bibinfo{year}{2023}),
  \bibinfo{numpages}{26}~pages.
\newblock
\urldef\tempurl%
\url{https://doi.org/10.1145/3611020}
\showDOI{\tempurl}


\bibitem[Powell et~al\mbox{.}(2006)]%
        {powell2006antecedents}
\bibfield{author}{\bibinfo{person}{Anne Powell}, \bibinfo{person}{John Galvin},
  {and} \bibinfo{person}{Gabriele Piccoli}.} \bibinfo{year}{2006}\natexlab{}.
\newblock \showarticletitle{Antecedents to team member commitment from near and
  far: A comparison between collocated and virtual teams}.
\newblock \bibinfo{journal}{\emph{IT \& People}}  \bibinfo{volume}{19}
  (\bibinfo{date}{10} \bibinfo{year}{2006}), \bibinfo{pages}{299--322}.
\newblock
\urldef\tempurl%
\url{https://doi.org/10.1108/09593840610718018}
\showDOI{\tempurl}


\bibitem[Ramun(2023)]%
        {ramun2023}
\bibfield{author}{\bibinfo{person}{Sadakshi~Kalyan Ramun}.}
  \bibinfo{year}{2023}\natexlab{}.
\newblock \bibinfo{title}{Riot games to scrap the Bait Ping in league of
  legends after misuse}.
\newblock
\newblock
\urldef\tempurl%
\url{https://afkgaming.com/esports/news/riot-games-to-scrap-the-bait-ping-in-league-of-legends-after-misuse}
\showURL{%
\tempurl}


\bibitem[Roberts et~al\mbox{.}(2014)]%
        {roberts2014}
\bibfield{author}{\bibinfo{person}{Nicole~K Roberts}, \bibinfo{person}{Reed~G
  Williams}, \bibinfo{person}{Cathy~J Schwind}, \bibinfo{person}{John~A
  Sutyak}, \bibinfo{person}{Christopher McDowell}, \bibinfo{person}{David
  Griffen}, \bibinfo{person}{Jarrod Wall}, \bibinfo{person}{Hilary Sanfey},
  \bibinfo{person}{Audra Chestnut}, \bibinfo{person}{Andreas~H Meier},
  {et~al\mbox{.}}} \bibinfo{year}{2014}\natexlab{}.
\newblock \showarticletitle{The impact of brief team communication, leadership
  and team behavior training on ad hoc team performance in trauma care
  settings}.
\newblock \bibinfo{journal}{\emph{The American Journal of Surgery}}
  \bibinfo{volume}{207}, \bibinfo{number}{2} (\bibinfo{year}{2014}),
  \bibinfo{pages}{170--178}.
\newblock


\bibitem[Seng{\"u}n et~al\mbox{.}(2019)]%
        {sengun2019exploring}
\bibfield{author}{\bibinfo{person}{Sercan Seng{\"u}n}, \bibinfo{person}{Joni
  Salminen}, \bibinfo{person}{Peter Mawhorter}, \bibinfo{person}{Soon-gyo
  Jung}, {and} \bibinfo{person}{Bernard Jansen}.}
  \bibinfo{year}{2019}\natexlab{}.
\newblock \showarticletitle{Exploring the relationship between game content and
  culture-based toxicity: a case study of league of legends and MENA players}.
  In \bibinfo{booktitle}{\emph{Proceedings of the 30th ACM Conference on
  Hypertext and Social Media}}. \bibinfo{pages}{87--95}.
\newblock


\bibitem[{\c{S}}eng{\"u}n et~al\mbox{.}(2022)]%
        {csengun2022players}
\bibfield{author}{\bibinfo{person}{Sercan {\c{S}}eng{\"u}n},
  \bibinfo{person}{Joao~M Santos}, \bibinfo{person}{Joni Salminen},
  \bibinfo{person}{Soon-gyo Jung}, {and} \bibinfo{person}{Bernard~J Jansen}.}
  \bibinfo{year}{2022}\natexlab{}.
\newblock \showarticletitle{Do players communicate differently depending on the
  champion played? Exploring the Proteus effect in League of Legends}.
\newblock \bibinfo{journal}{\emph{Technological Forecasting and Social Change}}
   \bibinfo{volume}{177} (\bibinfo{year}{2022}), \bibinfo{pages}{121556}.
\newblock


\bibitem[Stachowski et~al\mbox{.}(2009)]%
        {stachowski2009crises}
\bibfield{author}{\bibinfo{person}{Alicia Stachowski}, \bibinfo{person}{Seth
  Kaplan}, {and} \bibinfo{person}{Mary Waller}.}
  \bibinfo{year}{2009}\natexlab{}.
\newblock \showarticletitle{The Benefits of Flexible Team Interaction During
  Crises}.
\newblock \bibinfo{journal}{\emph{The Journal of applied psychology}}
  \bibinfo{volume}{94} (\bibinfo{date}{11} \bibinfo{year}{2009}),
  \bibinfo{pages}{1536--43}.
\newblock
\urldef\tempurl%
\url{https://doi.org/10.1037/a0016903}
\showDOI{\tempurl}


\bibitem[Strater et~al\mbox{.}(2008)]%
        {strater2008}
\bibfield{author}{\bibinfo{person}{Laura~D Strater}, \bibinfo{person}{Haydee~M
  Cuevas}, \bibinfo{person}{Erik~S Connors}, \bibinfo{person}{Diane~M
  Ungvarsky}, {and} \bibinfo{person}{Mica~R Endsley}.}
  \bibinfo{year}{2008}\natexlab{}.
\newblock \showarticletitle{Situation awareness and collaborative tool usage in
  ad hoc command and control teams}. In \bibinfo{booktitle}{\emph{Proceedings
  of the Human Factors and Ergonomics Society annual meeting}},
  Vol.~\bibinfo{volume}{52}. Sage Publications Sage CA: Los Angeles, CA,
  \bibinfo{pages}{468--472}.
\newblock


\bibitem[Tan et~al\mbox{.}(2021)]%
        {tan2021less}
\bibfield{author}{\bibinfo{person}{Evelyn Tan}, \bibinfo{person}{Alex Wade},
  \bibinfo{person}{Athanasios Kokkinakis}, \bibinfo{person}{Georgia Heyes},
  \bibinfo{person}{Simon~Peter Demediuk}, {and} \bibinfo{person}{Anders
  Drachen}.} \bibinfo{year}{2021}\natexlab{}.
\newblock \showarticletitle{Less is More: Analysing Communication in Teams of
  Strangers}.
\newblock  (\bibinfo{year}{2021}).
\newblock


\bibitem[Tan et~al\mbox{.}(2022)]%
        {tan2022}
\bibfield{author}{\bibinfo{person}{Evelyn T~S Tan}, \bibinfo{person}{Katja
  Rogers}, \bibinfo{person}{Lennart~E. Nacke}, \bibinfo{person}{Anders
  Drachen}, {and} \bibinfo{person}{Alex Wade}.}
  \bibinfo{year}{2022}\natexlab{}.
\newblock \showarticletitle{Communication Sequences Indicate Team Cohesion: A
  Mixed-Methods Study of Ad Hoc League of Legends Teams}.
\newblock \bibinfo{journal}{\emph{Proc. ACM Hum.-Comput. Interact.}}
  \bibinfo{volume}{6}, \bibinfo{number}{CHI PLAY}, Article
  \bibinfo{articleno}{225} (\bibinfo{date}{oct} \bibinfo{year}{2022}),
  \bibinfo{numpages}{27}~pages.
\newblock
\urldef\tempurl%
\url{https://doi.org/10.1145/3549488}
\showDOI{\tempurl}


\bibitem[Tang et~al\mbox{.}(2012)]%
        {tang2012verbal}
\bibfield{author}{\bibinfo{person}{Anthony Tang}, \bibinfo{person}{Jonathan
  Massey}, \bibinfo{person}{Nelson Wong}, \bibinfo{person}{Derek Reilly}, {and}
  \bibinfo{person}{W.~Keith Edwards}.} \bibinfo{year}{2012}\natexlab{}.
\newblock \showarticletitle{Verbal coordination in first person shooter games}.
  In \bibinfo{booktitle}{\emph{Proceedings of the ACM 2012 Conference on
  Computer Supported Cooperative Work}} (Seattle, Washington, USA)
  \emph{(\bibinfo{series}{CSCW '12})}. \bibinfo{publisher}{Association for
  Computing Machinery}, \bibinfo{address}{New York, NY, USA},
  \bibinfo{pages}{579–582}.
\newblock
\showISBNx{9781450310864}
\urldef\tempurl%
\url{https://doi.org/10.1145/2145204.2145292}
\showDOI{\tempurl}


\bibitem[Taylor(2012)]%
        {taylor2012fps}
\bibfield{author}{\bibinfo{person}{Nick Taylor}.}
  \bibinfo{year}{2012}\natexlab{}.
\newblock \bibinfo{booktitle}{\emph{Guns, Grenades, and Grunts: First-Person
  Shooter Games} (\bibinfo{edition}{1st} ed.)}.
\newblock \bibinfo{publisher}{Continuum}.
\newblock
\showISBNx{144114224X}


\bibitem[the NPC(2024)]%
        {jarvis2024}
\bibfield{author}{\bibinfo{person}{Jarvis the NPC}.}
  \bibinfo{year}{2024}\natexlab{}.
\newblock
\newblock
\urldef\tempurl%
\url{https://www.zleague.gg/theportal/the-impact-of-the-fist-bump-emote-in-league-of-legends-a-community-discussion/#google_vignette}
\showURL{%
\tempurl}


\bibitem[Turbosmurfs(2024)]%
        {turbosmurfs}
\bibfield{author}{\bibinfo{person}{Turbosmurfs}.}
  \bibinfo{year}{2024}\natexlab{}.
\newblock \bibinfo{title}{League of Legends: Player Count and Statistics 2024}.
\newblock
\newblock
\urldef\tempurl%
\url{https://turbosmurfs.gg/article/league-of-legends-player-count-and-statistics}
\showURL{%
\tempurl}


\bibitem[Uitdewilligen and Waller(2018)]%
        {uitdewilligen2018crisis}
\bibfield{author}{\bibinfo{person}{Sjir Uitdewilligen} {and}
  \bibinfo{person}{Mary~J. Waller}.} \bibinfo{year}{2018}\natexlab{}.
\newblock \showarticletitle{Information sharing and decision-making in
  multidisciplinary crisis management teams}.
\newblock \bibinfo{journal}{\emph{Journal of Organizational Behavior}}
  \bibinfo{volume}{39}, \bibinfo{number}{6} (\bibinfo{year}{2018}),
  \bibinfo{pages}{731--748}.
\newblock
\urldef\tempurl%
\url{https://doi.org/10.1002/job.2301}
\showDOI{\tempurl}
\showeprint{https://onlinelibrary.wiley.com/doi/pdf/10.1002/job.2301}


\bibitem[Veltri et~al\mbox{.}(2014)]%
        {veltri2014gender}
\bibfield{author}{\bibinfo{person}{Natasha Veltri}, \bibinfo{person}{Hanna
  Krasnova}, \bibinfo{person}{Annika Baumann}, {and} \bibinfo{person}{Neena
  Kalayamthanam}.} \bibinfo{year}{2014}\natexlab{}.
\newblock \showarticletitle{Gender differences in online gaming: A literature
  review}.
\newblock  (\bibinfo{year}{2014}).
\newblock


\bibitem[Vinella et~al\mbox{.}(2022)]%
        {vinella2022personality}
\bibfield{author}{\bibinfo{person}{Federica Vinella}, \bibinfo{person}{Chinasa
  Odo}, \bibinfo{person}{I. Lykourentzou}, {and} \bibinfo{person}{Judith
  Masthoff}.} \bibinfo{year}{2022}\natexlab{}.
\newblock \showarticletitle{How Personality and Communication Patterns Affect
  Online ad-hoc Teams Under Pressure}.
\newblock \bibinfo{journal}{\emph{Frontiers in Artificial Intelligence}}
  \bibinfo{volume}{5} (\bibinfo{date}{05} \bibinfo{year}{2022}),
  \bibinfo{pages}{818491}.
\newblock
\urldef\tempurl%
\url{https://doi.org/10.3389/frai.2022.818491}
\showDOI{\tempurl}


\bibitem[White et~al\mbox{.}(2020)]%
        {white2020covid}
\bibfield{author}{\bibinfo{person}{Bobbie White}, \bibinfo{person}{Justin
  Johnson}, \bibinfo{person}{Alejandro Arroliga}, {and} \bibinfo{person}{Glen
  Couchman}.} \bibinfo{year}{2020}\natexlab{}.
\newblock \showarticletitle{Ad hoc teams and telemedicine during COVID-19}.
\newblock \bibinfo{journal}{\emph{Proceedings (Baylor University. Medical
  Center)}}  \bibinfo{volume}{33} (\bibinfo{date}{09} \bibinfo{year}{2020}),
  \bibinfo{pages}{696--698}.
\newblock
\urldef\tempurl%
\url{https://doi.org/10.1080/08998280.2020.1809758}
\showDOI{\tempurl}


\bibitem[White et~al\mbox{.}(2018)]%
        {white2018}
\bibfield{author}{\bibinfo{person}{Bobbie Ann~A. White},
  \bibinfo{person}{Angela Eklund}, \bibinfo{person}{Tresa McNeal},
  \bibinfo{person}{Angie Hochhalter}, {and} \bibinfo{person}{Alejandro~C.
  Arroliga}.} \bibinfo{year}{2018}\natexlab{}.
\newblock \showarticletitle{Facilitators and barriers to ad hoc team
  performance}.
\newblock \bibinfo{journal}{\emph{Baylor University Medical Center
  Proceedings}} \bibinfo{volume}{31}, \bibinfo{number}{3}
  (\bibinfo{year}{2018}), \bibinfo{pages}{380--384}.
\newblock
\urldef\tempurl%
\url{https://doi.org/10.1080/08998280.2018.1457879}
\showDOI{\tempurl}


\bibitem[Wildman et~al\mbox{.}(2012)]%
        {wildman2012trust}
\bibfield{author}{\bibinfo{person}{Jessica Wildman}, \bibinfo{person}{Marissa
  Shuffler}, \bibinfo{person}{Elizabeth Lazzara}, \bibinfo{person}{Stephen
  Fiore}, \bibinfo{person}{Shawn Burke}, \bibinfo{person}{Eduardo Salas}, {and}
  \bibinfo{person}{Sena Garven}.} \bibinfo{year}{2012}\natexlab{}.
\newblock \showarticletitle{Trust Development in Swift Starting Action Teams: A
  Multilevel Framework}.
\newblock \bibinfo{journal}{\emph{Group \& Organization Management - GROUP
  ORGAN MANAGE}}  \bibinfo{volume}{37} (\bibinfo{date}{04}
  \bibinfo{year}{2012}), \bibinfo{pages}{137--170}.
\newblock
\urldef\tempurl%
\url{https://doi.org/10.1177/1059601111434202}
\showDOI{\tempurl}


\bibitem[Zheng and Farzan(2023)]%
        {zheng2023}
\bibfield{author}{\bibinfo{person}{Keyang Zheng} {and} \bibinfo{person}{Rosta
  Farzan}.} \bibinfo{year}{2023}\natexlab{}.
\newblock \showarticletitle{Understanding Player’s Gesture-Based
  Communicative Behavior in MOBA Games}.
\newblock \bibinfo{journal}{\emph{Proc. ACM Hum.-Comput. Interact.}}
  \bibinfo{volume}{7}, \bibinfo{number}{CHI PLAY}, Article
  \bibinfo{articleno}{415} (\bibinfo{date}{oct} \bibinfo{year}{2023}),
  \bibinfo{numpages}{23}~pages.
\newblock
\urldef\tempurl%
\url{https://doi.org/10.1145/3611061}
\showDOI{\tempurl}


\bibitem[Zijlstra et~al\mbox{.}(2012)]%
        {zijlstra2012interaction}
\bibfield{author}{\bibinfo{person}{Fred Zijlstra}, \bibinfo{person}{Mary
  Waller}, {and} \bibinfo{person}{Sybil Phillips}.}
  \bibinfo{year}{2012}\natexlab{}.
\newblock \showarticletitle{Setting the tone: Early interaction patterns in
  swift-starting teams as a predictor of effectiveness}.
\newblock \bibinfo{journal}{\emph{European Journal of Work and Organizational
  Psychology}}  \bibinfo{volume}{21} (\bibinfo{date}{10} \bibinfo{year}{2012}).
\newblock
\urldef\tempurl%
\url{https://doi.org/10.1080/1359432X.2012.690399}
\showDOI{\tempurl}


\end{thebibliography}

\appendix
\clearpage
\section{Codebook}
We present the full codebook of the study data below.

\label{appendix}
\begin{table}[!htbp]
\centering
\renewcommand{\arraystretch}{0.85}
\parbox{\textwidth}{\caption{Codebook from Thematic Analysis of Experiment Data}}
\vspace{-3pt}
{\begin{tabular}{ll|l}
\toprule
\multicolumn{1}{l|}{\textbf{Research Question}}                                                                                                                                                          & \textbf{Category}                                                                                           & \textbf{Code}                                                       \\ \hline
\multicolumn{1}{l|}{\multirow{20}{*}{\begin{tabular}[c]{@{}l@{}}When and why do players \\ engage in in-game \\ communication in \\ \textit{League of Legends}?\end{tabular}}}                                    & \multirow{7}{*}{\begin{tabular}[c]{@{}l@{}}Communication \\ Engagement Types\end{tabular}}                  & Encouraging the teammate                                            \\ \cline{3-3} 
\multicolumn{1}{l|}{}                                                                                                                                                                                    &                                                                                                             & Praising the teammate                                               \\ \cline{3-3} 
\multicolumn{1}{l|}{}                                                                                                                                                                                    &                                                                                                             & Apologizing for a bad play                                          \\ \cline{3-3} 
\multicolumn{1}{l|}{}                                                                                                                                                                                    &                                                                                                             & Consoling the teammate                                              \\ \cline{3-3} 
\multicolumn{1}{l|}{}                                                                                                                                                                                    &                                                                                                             & Asking for a specific play                                          \\ \cline{3-3} 
\multicolumn{1}{l|}{}                                                                                                                                                                                    &                                                                                                             & Criticizing or expressing disapproval                               \\ \cline{3-3} 
\multicolumn{1}{l|}{}                                                                                                                                                                                    &                                                                                                             & Providing feedback                                                  \\ \cline{2-3} 
\multicolumn{1}{l|}{}                                                                                                                                                                                    & \multirow{6}{*}{\begin{tabular}[c]{@{}l@{}}Communication \\ Engagement Triggers\end{tabular}}               & Teammate performs badly                                             \\ \cline{3-3} 
\multicolumn{1}{l|}{}                                                                                                                                                                                    &                                                                                                             & Teammate makes great play                                           \\ \cline{3-3} 
\multicolumn{1}{l|}{}                                                                                                                                                                                    &                                                                                                             & Player performs badly                                               \\ \cline{3-3} 
\multicolumn{1}{l|}{}                                                                                                                                                                                    &                                                                                                             & Teammate starts conflict with the player                            \\ \cline{3-3} 
\multicolumn{1}{l|}{}                                                                                                                                                                                    &                                                                                                             & Teammate specifically calls out the player                          \\ \cline{3-3} 
\multicolumn{1}{l|}{}                                                                                                                                                                                    &                                                                                                             & No specific trigger                                                 \\ \cline{2-3} 
\multicolumn{1}{l|}{}                                                                                                                                                                                    & \multirow{7}{*}{\begin{tabular}[c]{@{}l@{}}Reasons Behind \\ Not Engaging in \\ Communication\end{tabular}} & Not enough time                                                     \\ \cline{3-3} 
\multicolumn{1}{l|}{}                                                                                                                                                                                    &                                                                                                             & Does not believe that the communciation will have impact            \\ \cline{3-3} 
\multicolumn{1}{l|}{}                                                                                                                                                                                    &                                                                                                             & Does not want to cause friction or trigger offensive players        \\ \cline{3-3} 
\multicolumn{1}{l|}{}                                                                                                                                                                                    &                                                                                                             & Didn't realize the communication was sent or misinterpreted it      \\ \cline{3-3} 
\multicolumn{1}{l|}{}                                                                                                                                                                                    &                                                                                                             & Believes that the game outcome is already determined                \\ \cline{3-3} 
\multicolumn{1}{l|}{}                                                                                                                                                                                    &                                                                                                             & Sufficient communication through other medium                       \\ \cline{3-3} 
\multicolumn{1}{l|}{}                                                                                                                                                                                    &                                                                                                             & Makes it difficult to focus on the game                             \\ \hline
\multicolumn{1}{l|}{\multirow{24}{*}{\begin{tabular}[c]{@{}l@{}}How do players assess and \\ react to in-game team \\ communication in real time?\end{tabular}}}                                         & \multirow{4}{*}{Communication Cost}                                                                         & Communication speed                                                 \\ \cline{3-3} 
\multicolumn{1}{l|}{}                                                                                                                                                                                    &                                                                                                             & Ability to convey the granularity of desired communication          \\ \cline{3-3} 
\multicolumn{1}{l|}{}                                                                                                                                                                                    &                                                                                                             & Possibility of misuse or abuse                                      \\ \cline{3-3} 
\multicolumn{1}{l|}{}                                                                                                                                                                                    &                                                                                                             & Distraction to the gameplay                                         \\ \cline{2-3} 
\multicolumn{1}{l|}{}                                                                                                                                                                                    & \multirow{3}{*}{\begin{tabular}[c]{@{}l@{}}Relevance and \\ Responsiveness\end{tabular}}                    & Evaluation of whether the situation is ongoing                      \\ \cline{3-3} 
\multicolumn{1}{l|}{}                                                                                                                                                                                    &                                                                                                             & Showing through action                                              \\ \cline{3-3} 
\multicolumn{1}{l|}{}                                                                                                                                                                                    &                                                                                                             & Interpreting other players' responses                               \\ \cline{2-3} 
\multicolumn{1}{l|}{}                                                                                                                                                                                    & \multirow{3}{*}{Psychological Safety}                                                                       & Cause of stress                                                     \\ \cline{3-3} 
\multicolumn{1}{l|}{}                                                                                                                                                                                    &                                                                                                             & Reduction in mental fortitude and will to play                      \\ \cline{3-3} 
\multicolumn{1}{l|}{}                                                                                                                                                                                    &                                                                                                             & Improvement to players' mood                                        \\ \cline{2-3} 
\multicolumn{1}{l|}{}                                                                                                                                                                                    & \multirow{5}{*}{Reducing Friction}                                                                          & Increased likelihood for team loss                                  \\ \cline{3-3} 
\multicolumn{1}{l|}{}                                                                                                                                                                                    &                                                                                                             & Cause of conflict within the team                                   \\ \cline{3-3} 
\multicolumn{1}{l|}{}                                                                                                                                                                                    &                                                                                                             & Usage bias towards negative communication                           \\ \cline{3-3} 
\multicolumn{1}{l|}{}                                                                                                                                                                                    &                                                                                                             & Concealing gender identity to prevent harassment                    \\ \cline{3-3} 
\multicolumn{1}{l|}{}                                                                                                                                                                                    &                                                                                                             & Valuing player identity and preference                              \\ \cline{2-3} 
\multicolumn{1}{l|}{}                                                                                                                                                                                    & \multirow{4}{*}{Communication Hierarchy}                                                                    & Player playing well should lead communication                       \\ \cline{3-3} 
\multicolumn{1}{l|}{}                                                                                                                                                                                    &                                                                                                             & Player playing badly should not communicate or would not be trusted \\ \cline{3-3} 
\multicolumn{1}{l|}{}                                                                                                                                                                                    &                                                                                                             & Communication when player's influence is low on the game            \\ \cline{3-3} 
\multicolumn{1}{l|}{}                                                                                                                                                                                    &                                                                                                             & Communication when player's influence is high in the game           \\ \cline{2-3} 
\multicolumn{1}{l|}{}                                                                                                                                                                                    & \multirow{2}{*}{Norms and Habits}                                                                           & Habitual usage without specific intent                              \\ \cline{3-3} 
\multicolumn{1}{l|}{}                                                                                                                                                                                    &                                                                                                             & Not used to using the communication channel                         \\ \cline{2-3} 
\multicolumn{1}{l|}{}                                                                                                                                                                                    & \multirow{3}{*}{Reaction}                                                                                   & Uses information from communication to inform future decisions      \\ \cline{3-3} 
\multicolumn{1}{l|}{}                                                                                                                                                                                    &                                                                                                             & Follows calls perceived to be good                                  \\ \cline{3-3} 
\multicolumn{1}{l|}{}                                                                                                                                                                                    &                                                                                                             & Muting communication channels                                       \\ \hline
\multicolumn{1}{l|}{\multirow{7}{*}{\begin{tabular}[c]{@{}l@{}}How does the player's \\ experience with in-game \\ communication processes \\ hape their perception \\ towards teammates?\end{tabular}}} & \multirow{4}{*}{Trust in the Teammate}                                                                      & Bad starts in communication as negative portend                     \\ \cline{3-3} 
\multicolumn{1}{l|}{}                                                                                                                                                                                    &                                                                                                             & Distrust and dislike of talkative teammates                         \\ \cline{3-3} 
\multicolumn{1}{l|}{}                                                                                                                                                                                    &                                                                                                             & Anxiousness of teammates turning hostile                            \\ \cline{3-3} 
\multicolumn{1}{l|}{}                                                                                                                                                                                    &                                                                                                             & Importance of strong mental fortitude and will to play              \\ \cline{2-3} 
\multicolumn{1}{l|}{}                                                                                                                                                                                    & \multirow{3}{*}{Player Commitment Fortitude}                                                                & Leading the team and calling shots                                  \\ \cline{3-3} 
\multicolumn{1}{l|}{}                                                                                                                                                                                    &                                                                                                             & Communicating in advance                                            \\ \cline{3-3} 
\multicolumn{1}{l|}{}                                                                                                                                                                                    &                                                                                                             & Positive attitude in communication                                  \\ \hline
\multicolumn{2}{l|}{\multirow{4}{*}{Other}}                                                                                                                                                                                                                                                                            & Reason for playing ranked League of Legends                         \\ \cline{3-3} 
\multicolumn{2}{l|}{}                                                                                                                                                                                                                                                                                                  & Differences in communication when playing with friends              \\ \cline{3-3} 
\multicolumn{2}{l|}{}                                                                                                                                                                                                                                                                                                  & Benefits and drawbacks of voice-based communication                 \\ \cline{3-3} 
\multicolumn{2}{l|}{}                                                                                                                                                                                                                                                                                                  & Benefits and drawbacks of communication with the enemy team        \\
\bottomrule
\end{tabular}}
\label{tab:codebook}
\Description{
    Table showing the codebook resulting from our thematic analysis, which lists out research questions, categories, and codes.}
\end{table}

\clearpage
\section{Interview Questions}
\label{appendix2}
We list out the full interview questions of the study below.
\\
\\
\textbf{Game History}
\begin{itemize}
\item How long have you been playing \textit{League of Legends}?
\item What’s the primary reason you play \textit{League of Legends}?
\end{itemize}

\textbf{Pre-game Stage (Ban/Pick)}
\begin{itemize}
\item Why/why are you not communicating on the chat?
\item(If teammate communicated) What do you think about what the teammate said?
\item Why did you/Why did you not reply to them?
\item What do you think about your team/teammates?
\end{itemize}

\textbf{In-Game Stage}

\textit{Chat Usage}
\begin{itemize}
\item Why did you use say [message]?
\item What was the purpose of using [communication channel]?
\item Who was your communication for?
\item What did you mean when you said [message]?
\item (If a teammate sent a message) What do you think the message meant? How confident are you in your answer?
\item (If a teammate sent a message) Who do you think the message was meant for?
\item (If a teammate sent a message) Why do you think they sent the message?
\item (If a teammate sent a message) What do you think about the player after reading the message?
\item (If a teammate sent a message) How do you feel about the communication?
\item You don’t seem to use/it seems like you frequently use chat. Why is this so?
\end{itemize}
 
\textit{Ping, Emote Usage}
\begin{itemize}
\item Why did you use the ping/emote?
\item What were you trying to convey/emote?
\item Who were you targeting?
\item Do you think the target understood what you were trying to convey?
\item (If a teammate pings/emotes) What do you think the ping/emote signified? How confident are you in your answer?
\item (If a teammate pings/emotes) How do you feel about the communication?
\item You don’t seem to use/it seems like you frequently use pings/emotes. Why is this so?
\end{itemize}

\textit{Vote Usage}
\begin{itemize}
\item (If the player began the vote) Why did you begin the vote?
\item (If a teammate began the vote) Why did you vote [yes or no]/not vote?
\item How do you feel about the teammates’ (lack of) response to the vote?
\item You don’t seem to vote/it seems like you always vote. Why is this so?
\end{itemize}

\textit{Situational}
\begin{itemize}
\item You just [communication behavior]. What was the purpose?
\item Why did you mute [player, communication tool]?
\item Why did you stop [communication attempt]?
\item Previously, you said you did not do [communication behavior]. Why did you do it this time? 
\end{itemize}

\textbf{Post-Game Stage}
\begin{itemize}
\item It seems that you tended to [communication behavior] throughout the game. Why is this so?
\item How did the teammate’s communication behavior affect your perception? If it changed throughout the game, how and why did it change?
\item If you communication behavior changed throughout the game, how and why did it change?
\item How much do you trust your teammates?
\item How do you manage hostile communication?
\item What do you think about team communication during this game?
\item Do you have experiences with another player specifying or asking about your gender? If so, how did you handle those cases?
\item How do you think the other player guessed your gender?
\item How do you feel about the other players’ comments based on your gender identity?
\item What do you think about implementing voice chats in \textit{League of Legends} in Solo Rank mode?
\item What do you think about the removal of all chat in the \textit{League of Legends}?
\item What other communication features do you wish \textit{League of Legends} had?
\item What are some ways that communication in \textit{League of Legends} could be improved?
\end{itemize}

\end{document}